\documentclass[11pt,a4paper]{article}
\usepackage{arxiv}
\usepackage[sort,comma,authoryear,round]{natbib}

\usepackage{graphicx}
\usepackage{amsmath, bm, amssymb, float, comment, amsthm}
\usepackage{enumitem}
\usepackage{verbatim}
\usepackage{rotating}
\usepackage{setspace}
\usepackage[nottoc]{tocbibind}
\usepackage[usenames, dvipsnames]{xcolor}
\usepackage{soul}

\definecolor{PineGreen}{rgb}{0, 0, 0}

\usepackage{hyperref}
\hypersetup{
    colorlinks=true,
    linkcolor=blue,
    filecolor=magenta,      
    urlcolor=cyan,
    citecolor=blue,
    }

\theoremstyle{definition}
\newtheorem{remark}{Remark}
\newcommand{\Hline}{\hline}

\newcommand{\boldxi}{\bm{\xi}}
\newcommand{\boldmu}{\bm{\mu}}

\newcommand{\boldbeta}{\bm{\beta}}

\newcommand{\boldzeta}{\bm{\zeta}}
\newcommand{\boldSigma}{\bm{\varSigma}}
\newcommand{\boldepsilon}{\bm{\epsilon}}
\newcommand{\boldeta}{\bm{\eta}}

\newcommand{\boldomega}{\bm{\omega}}
\newcommand{\boldtheta}{\bm{\theta}}

\newcommand{\boldLambda}{\bm{\Lambda}}
\newcommand{\boldXi}{\bm{\Xi}}
\newcommand{\sigmaeps}{\sigma_\epsilon}

\title{Regularized Principal Spline Functions to Mitigate Spatial Confounding}

\author{
 	Carlo Zaccardi \\
 	Department of Economics\\
	University G. d'Annunzio of Chieti-Pescara\\
	Pescara, Italy \\
  \texttt{carlo.zaccardi@unich.it} \\
  \And
	Pasquale Valentini \\
 	Department of Economics\\
	University G. d'Annunzio of Chieti-Pescara\\
	Pescara, Italy \\
  \And
	Luigi Ippoliti \\
 	Department of Economics\\
	University G. d'Annunzio of Chieti-Pescara\\
	Pescara, Italy \\
	\And
	Alexandra M. Schmidt \\
  Department of Epidemiology, Biostatistics and Occupational Health\\
  McGill University\\
  Montréal, Canada \\
}

\begin{document}

\maketitle


\begin{abstract}
	This paper proposes a new approach to address the problem of unmeasured confounding in spatial designs. Spatial confounding occurs when some confounding variables are unobserved and not included in the model, leading to distorted inferential results about the effect of an exposure on an outcome.
	We show the relationship existing between the confounding bias of a non-spatial model and that of a semi-parametric model that includes a basis matrix to represent the unmeasured confounder conditional on the exposure. This relationship holds for any basis expansion, however it is shown that using the semi-parametric approach guarantees a reduction in the confounding bias only under certain circumstances, which are related to the spatial structures of the exposure and the unmeasured confounder, the type of basis expansion utilized, and the regularization mechanism.
	To adjust for spatial confounding, and therefore try to recover the effect of interest, we propose a Bayesian semi-parametric regression model, where an expansion matrix of principal spline basis functions is used to approximate the unobserved factor, and spike-and-slab priors are imposed on the respective expansion coefficients in order to select the most important bases. From the results of an extensive simulation study, we conclude that our proposal is able to reduce the confounding bias more than competing approaches, and it also seems more robust to bias amplification.
\end{abstract}

%


\section{Introduction}
\label{sec:1}

In many observational studies, an important statistical problem lies in inferring the effects of one or more predictors, also known as treatments or exposures, on a response variable, while accounting for confounders, i.e. factors associated with both exposures and response \citep[see, e.g.,][]{lawson2018bayesian}. However, adjusting for relevant confounders can be challenging if these variables are unknown. This limitation poses a significant hurdle as standard estimators are often biased and inconsistent. Hence, a major concern in estimating exposure effects is how to account for unmeasured confounders. In spatially structured data, this challenge is known as spatial confounding, first identified by \citet{clayton1993spatial}.

In a regression framework, various methodological approaches have been advanced to mitigate the bias stemming from spatial confounding. For recent literature reviews see, for example, \citet{reich2021review}, \citet{khan2023re}, and \citet{urdangarin2023evaluating}. To adjust for unobserved covariates and improve model fitting, it is possible to include a spatially correlated random effect in the model. Nevertheless, research has demonstrated that this strategy does not necessarily alleviate the unmeasured confounding bias \citep{clayton1993spatial,paciorek2010importance,page2017estimation}. In particular, \citet{hodges2010adding} demonstrated that this approach can actually increase bias relative to fitting a non-spatial model, due to the collinearity between the spatial term and the exposure.
As a result, \citet{reich2006effects} and \citet{hodges2010adding} propose the restricted spatial regression (RSR) model, wherein a spatial random effect (SRE) is included but is restricted to the orthogonal complement of the space spanned by other observed covariates. Various revisions and enhancements to RSR have been proposed in subsequent studies \citep{azevedo2021mspock,dupont2022spatialplus, hanks2015restricted, hughes2013dimension, nobre2021effects, prates2019alleviating}. However, research has shown that approaches based on RSR are not effective in mitigating spatial confounding \citep{khan2022restricted, zimmerman2022deconfounding}. 

In parallel with the development of RSR, \citet{paciorek2010importance} explored further scenarios where fitting a spatial model may lead to changes in bias. In the context of Gaussian processes, the author provides an inequality condition for bias reduction (with respect to a non-spatial model that ignores confounding), where the variability of the exposure must be confined within a smaller spatial range compared to that of the unmeasured confounder. Reversing this inequality, confounding bias becomes amplified. Besides, the non-spatial variation in the exposure is also found to be impactful on the bias \citep{bobb2022accounting}. These results, confirmed also by \citet{page2017estimation}, guided the development of several solutions, based on different reparametrizations of a spatial linear model \citep{bobb2022accounting, dupont2022spatialplus, guan2023spectral, keller2020selecting, marques2022mitigating, prates2015transformed, thaden2018structural,urdangarin2024simplified}.

A popular approach involves modelling the random effect component as a smooth function of spatial locations using spline basis functions \citep{dupont2022spatialplus, guan2023spectral, keller2020selecting, paciorek2010importance}. This is motivated by the versatility of splines to address spatial scale and smoothness issues in both measured and unmeasured variables. However, given the potential for increased bias, a critical question in data analysis is to determine the ``optimal'' adjustment. 
Proposed approaches involve utilizing information criteria to determine the number of basis functions \citep{guan2023spectral, keller2020selecting}, or employing generalized cross-validation to select a smoothing parameter \citep{dupont2022spatialplus}.

Central to our analysis is the formulation of the spatial model as a reduced-rank Bayesian regression model, where we approximate the unmeasured spatial confounder, conditioned on the exposure, using an expansion matrix of \textit{principal kriging functions}. Notably, these basis functions are data-adaptive and can be organized from coarse to fine scales of spatial variation \citep{fontanella2013functional, kent2001functional, mardia1998kriged}. For a specific parametrization, we demonstrate that the principal kriging functions lead to principal thin-plate splines functions, thus establishing a connection between splines and kriging. Within the Bayesian paradigm, we then impose \textit{non-local spike-and-slab priors} on the basis coefficients \citep{johnson2012bayesian,rossell2017nonlocal}. Unlike the sequential selection of bases obtained via information criteria, the proposed methodology conducts variable and model selection simultaneously, thereby avoiding issues associated with truncating the basis matrix or selecting a smoothing parameter.

We envision spatial confounding as a ``data generation" problem, that is the problematic relationship is between the processes generating exposure and unmeasured confounder \citep{khan2023re}. To investigate the mechanism of spatial confounding within the proposed reduced-rank model, we provide a theoretical expression of the bias of the \textit{adjusted ordinary least squares} estimator which clearly shows how the spatial scale of the involved spatial processes and the selection of the basis functions can either mitigate or exacerbate the bias.

The paper is structured as follows. Section \ref{sec:2} reviews the theoretical framework. Section \ref{sec:3} introduces the spatial linear model and its representation based on principal kriging and spline functions. Section \ref{sec:why} discusses the circumstances under which confounding bias can be alleviated. Section \ref{sec:model} discusses the proposed Bayesian model regularization approach. Section \ref{sec:simulations} illustrates the results from a wide range of simulated scenarios, where the proposed approach is also compared to most of the recently published methods. Section \ref{sec:application} applies these methods to real data concerning tropospheric ozone, Section \ref{sec:discussion} concludes with a discussion.

\section{Preliminaries}
\label{sec:2}


Consider the usual Gaussian spatial linear model \citep{cressie1993spatial}:
\begin{equation} \label{eq:main}
	Y(\mathbf{s}_i) = \beta_0 + \beta_x X(\mathbf{s}_i) + W(\mathbf{s}_i)  + \epsilon_y(\mathbf{s}_i), \qquad \epsilon_y(\mathbf{s}_i) \stackrel{{ iid }}{\sim} N(0,\sigmaeps^2)\,,
\end{equation}
where $Y(\mathbf{s}_i)$ is the outcome variable, $ X(\mathbf{s}_i) $ denotes the exposure with unknown effect, $\beta_x$, and $W(\mathbf{s}_i)$ is an unobserved spatially-varying random field. All variables are defined over a continuous spatial domain, $\mathcal{S} \subseteq \mathbb{R}^2$, and $Y(\cdot)$ and $X(\cdot)$ are observed at a finite set of locations, $ \lbrace \mathbf{s}_1, \dots, \mathbf{s}_n \rbrace $, for any $ n \in \mathbb{N}\backslash\{0\} $, where $\mathbf{s}_i = (\mathbf{s}_i[1], \mathbf{s}_i[2]) \in \mathcal{S}$ is a location, with $\mathbf{s}[1]$ and $\mathbf{s}[2]$ denoting its easting and northing coordinates. Accordingly, if $\mathbf{Y}$, $\mathbf{X}$, $\mathbf{W}$ and $\boldepsilon$ are $n$-dimensional random vectors, Equation (\ref{eq:main}) can be rewritten in matrix formulation as
\begin{equation} \label{eq:spatial-linear-model-matrix}
	\mathbf{Y} = \Tilde{\mathbf{X}} \boldbeta + \Tilde{\mathbf{W}}\,,
\end{equation}
where $\Tilde{\mathbf{W}} = \mathbf{W} + \boldepsilon$ is the spatial error term, $ \Tilde{\mathbf{X}} = [\mathbf{1}_n \ \mathbf{X}] $, with $\mathbf{1}_n$ an $n$-dimensional vector of ones, and $\boldbeta = (\beta_0, \beta_x)'$. A further usual assumption \citep{banerjee2014hierarchical,cressie1993spatial} is that $X(\cdot)$ and $W(\cdot)$ are independent. 
%
Provided that the variance parameters are known, the generalized least squares (GLS) estimator of $\boldbeta$ is $
	\widehat{\boldbeta}_{GLS}=\left(\Tilde{\bf X}^{\prime} \boldSigma_{\widetilde{w}}^{-1}\Tilde{\bf X}\right)^{-1} \Tilde{\bf X}^{\prime} \boldSigma_{\widetilde{w}}^{-1}{\bf Y}
$, which is unbiased and the most efficient within the class of linear estimators.

In the presence of spatial confounding, however, the regressor $\mathbf{X}$ is not independent of the unobserved random field $\mathbf{W}$ \citep{clayton1993spatial}. Therefore, $\widehat{\boldbeta}_{GLS}$ is no longer an unbiased estimator of $\boldbeta$ and, in a regression setting where $\mathbf{Y}|\mathbf{X}$ is of principal interest, the outcome $\mathbf{Y}$ should be marginalized over $\mathbf{W}|\mathbf{X}$ \citep{paciorek2010importance,page2017estimation}; see also \citet{guan2023spectral} for a discussion in the spectral domain.

To better understand the problem, assume that the joint distribution of $\mathbf{X}$ and $\mathbf{W}$ is
\begin{equation} \label{eq:jointNormal}
	\left[\begin{array}{c}
		\mathbf{X} \\
		\mathbf{W}
	\end{array}\right] \sim N\left(\left[\begin{array}{c}
		\boldmu_x \\
		\boldmu_w
	\end{array}\right] , \left[\begin{array}{cc}
		\boldSigma_{x} & \boldSigma_{xw} \\
		\boldSigma_{wx} & \boldSigma_{w}
	\end{array}\right]\right),
\end{equation}
from which it also follows that:
\begin{itemize}
	\item $\mathbf{W}|\mathbf{X=x} \sim N (\boldmu_{w|x}, \boldSigma_{w|x}) $, where $
		\boldmu_{w|x} = \mbox{E} \left[\mathbf{W}|\mathbf{X=x}\right] = \boldmu_w + \boldSigma_{wx} \boldSigma_{x}^{-1} (\mathbf{x} - \boldmu_x) $, and $
		\boldSigma_{w|x} = \mbox{Var} \left[\mathbf{W}|\mathbf{X=x}\right] = \boldSigma_{w} - \boldSigma_{wx} \boldSigma_{x}^{-1} \boldSigma_{xw} $.
	\item $\mathbf{Y}|\mathbf{X=x} \sim N (\boldmu_{y|x}, \boldSigma_{y|x}) $, where $ \boldmu_{y|x} = E \left[\mathbf{Y}|\mathbf{X=x}\right] = \beta_0 \mathbf{1}_n + \beta_x \mathbf{x} + \boldmu_{w|x} $, $ \boldSigma_{y|x} = Var \left[\mathbf{Y}|\mathbf{X=x}\right] = \sigmaeps^2 \mathbf{I}_n + \boldSigma_{w|x} $, and $\mathbf{I}_n$ is the $n \times n$ identity matrix.
\end{itemize}

Then, following \citet{paciorek2010importance} and \citet{page2017estimation}, if all variance parameters are known and $\boldmu_x = \boldmu_w = \mathbf{0}$, the first moment of the GLS estimator is given by
\begin{align}
	E\left[\widehat{\boldbeta}_{GLS}|\mathbf{X=x}\right] &= \mathbf{H}_{GLS} \Tilde{\bf X} \boldbeta + \mathbf{H}_{GLS} E \left[\mathbf{W}|\mathbf{X=x}\right] = \boldbeta + \mathbf{H}_{GLS} \boldSigma_{wx} \boldSigma_{x}^{-1} \mathbf{x}\,, \label{eq:Ebeta_gls}
\end{align}
where $\mathbf{H}_{GLS} = \left(\Tilde{\mathbf{X}}' \boldSigma_{y|x}^{-1} \Tilde{\mathbf{X}} \right)^{-1} \Tilde{\mathbf{X}}' \boldSigma_{y|x}^{-1}$. Equation (\ref{eq:Ebeta_gls}) thus shows that the GLS estimator, $\widehat{\boldbeta}_{GLS}$, can be biased if $Cov(\mathbf{X, W}) \ne 0$, and that spatial confounding is a challenging and complex problem, since the spatial structure of $\bf W$ is unknown. Finally, under ordinary least squares (OLS) estimation, the expectation in Equation (\ref{eq:Ebeta_gls}) has the same form but with $\mathbf{H}_{OLS} = \left(\Tilde{\mathbf{X}}' \Tilde{\mathbf{X}} \right)^{-1} \Tilde{\mathbf{X}}'$ in place of $\mathbf{H}_{GLS}$.

There exist various model parametrizations to represent the correlation between $\mathbf{X}$ and $\mathbf{W}$, which will consequently shape Equation (\ref{eq:Ebeta_gls}). 
As in \citet{marques2022discussion}, \citet{marques2022mitigating}, \citet{paciorek2010importance}, \citet{page2017estimation}, and \citet{schmidt2022discussion}, it can be assumed that $\boldSigma_{xw}$ depend on the marginal covariance matrices. For $i,l=1,\dots,n$, consider an exponential correlation function such that $R(\| \mathbf{s}_i - \mathbf{s}_l \|;\phi) = \exp (- \phi^{-1} \| \mathbf{s}_i - \mathbf{s}_l \|)$, with $\phi > 0$ denoting the \textit{range} parameter. The covariance matrix of the exposure can thus be written as $\boldSigma_{x} = \sigma_{x}^2 \mathbf{R}_{\phi_x} $, where $\sigma_{x}>0$ and $ \mathbf{R}_{\phi_x} $ is a spatial correlation matrix with elements $\left( \mathbf{R}_{\phi_x} \right)_{il} = R(\| \mathbf{s}_i - \mathbf{s}_l \|;\phi_x)$. Assume also that $\mathbf{R}_{\phi_x}$ can be factorized as $ \mathbf{R}_{\phi_x} = \mathbf{R}_{\phi_x}^{1/2} \mathbf{R}_{\phi_x}^{1/2 \prime} $,	where $ \mathbf{R}_{\phi_x}^{1/2} $ may be obtained from a Cholesky or an eigenvalue decomposition. 
Similarly, we may write $\boldSigma_{w} = \sigma_{w}^2 \mathbf{R}_{\phi_w} = \sigma_{w}^2 \mathbf{R}_{\phi_w}^{1/2} \mathbf{R}_{\phi_w}^{1/2 \prime} $. Hence, the cross-covariance matrix is $\boldSigma_{xw} = \sigma_{xw} \mathbf{R}_{\phi_x}^{1/2} \mathbf{R}_{\phi_w}^{1/2\prime}$, where $ \sigma_{xw} = \delta \sigma_x \sigma_w $ and $ \delta \in (-1,1) $ is the linear correlation coefficient between $ \mathbf{X} $ and $ \mathbf{W} $. From this specification, if $\boldmu_w=\boldmu_x=\bf 0$, it is easy to show that $\boldmu_{w|x} =  \delta \frac{\sigma_w}{\sigma_x} \mathbf{R}_{\phi_w}^{1/2} \mathbf{R}_{\phi_x}^{-1/2} \mathbf{x}$ and $\boldSigma_{w|x} = \sigma_w^2 (1-\delta^2) \mathbf{R}_{\phi_w}$.
Also, the expected value of the GLS estimator is specified as follows:
\begin{gather*}
	E\left[\widehat{\boldbeta}_{GLS}|\mathbf{X=x}\right] = \boldbeta + \delta \frac{\sigma_w}{\sigma_x} \mathbf{H}_{GLS} \mathbf{R}_{\phi_w}^{1/2} \mathbf{R}_{\phi_x}^{-1/2} \mathbf{x}\,. 
\end{gather*}
The factorization approach explicitly shows that the structure of the bias of the GLS estimator depends on the magnitude of $\phi_w$ and $\phi_x$ and that it approaches zero if $\delta \rightarrow 0$ \citep{paciorek2010importance,page2017estimation}.

\section{A Low-Rank Smoother for Spatial Confounding} \label{sec:3}

As an alternative to Equation \eqref{eq:main}, a common approach in the literature for addressing spatial confounding involves recovering the coefficient of interest, $\beta_x$, by means of a semiparametric model that incorporates spatial splines to approximate unmeasured confounders. 
In our analysis, conditionally on the exposure $X(\mathbf{s})$, we work with a model having the following general form:
\begin{equation}\label{eq:semiparametric-model}
Y(\mathbf{s}_i)= f_x(\mathbf{s}_i) + \widetilde{\epsilon}_y (\mathbf{s}_i), \qquad \widetilde{\epsilon}_y (\mathbf{s}_i) \sim N(0,\widetilde{\sigma}_\epsilon^2)\,,
\end{equation}
where the effects of the exposure and the spatial process are represented by an unknown smooth function $f_x(\mathbf{s}_i)$ defined over the spatial domain of the data.

In general, two  different but related strategies can be used for estimating the function $f_x(\mathbf{s})$ given information  at a set of spatial sites. One is based on the theory of \textit{reproducing kernel Hilbert space} (RKHS), and involves minimizing a penalized function, while the other is based on optimal prediction in the stochastic process setting. The optimal smoothing problem in RKHS is detailed in Web Appendix A. In the following, we exploit their relationship to formulate a low-rank spatial model aimed at addressing spatial confounding.

\subsection{Smoothing and interpolation for stochastic processes} \label{subsec:smoothing}

While the connection between optimal smoothing in a separating RKHS framework and optimal prediction (kriging) for a stochastic process is well-established (see, \citealp{Wahba1990,cressie1993spatial,Kent1994}), it is valuable to explicitly outline these links within the context of spatial confounding, demonstrating how they lead to a reduced-rank representation of the conditional spatial model.

Let $\mathcal{H}$ be an RKHS with reproducing kernel $k(\mathbf{s}_i,\mathbf{s}_j)$ and suppose $\mathcal{H}$ can be written as an \textcolor{PineGreen}{orthogonal} direct sum $\mathcal{H}=\mathcal{H}_0 \bigoplus \mathcal{H}_1$ \citep{Wahba1990}, where $\mathcal{H}_0$ is a null space with basis $ \lbrace u_l(\mathbf{s}_i), l = 1,\dots, q \rbrace $. Let $Y(\mathbf{s}_i)$ be a stochastic process with mean and covariance structure
$$E \left[Y(\mathbf{s}_i) \right]=\sum_{l=1}^q g_l u_l(\mathbf{s}_i), \quad \quad  
Cov \bigl(Y(\mathbf{s}_i),Y(\mathbf{s}_j) \bigr)=k(\mathbf{s}_i,\mathbf{s}_j)\,.$$
Define $\mathbf{K}$  and $\mathbf{U}$ as $(n \times n)$ and  $(n \times q)$ matrices with elements $k_{ij}=k(\mathbf{s}_i,\mathbf{s}_j)$ and $u_{il}=u_{l}(\mathbf{s}_i)$, $i,j=1, \ldots, n$, $l=1, \ldots, q$. Then for any $\mathbf{s}$, the best linear unbiased (or kriging) predictor of $ Y(\mathbf{s})$ is identical to the value
of optimal smoothing function $f_x(\mathbf{s})$ in an RKHS \citep{Kent1994}, and both can be expressed in the form
\begin{equation*}
f_x(\mathbf{s}) = \Bigl(\mathbf{u}(\mathbf{s})' \mathbf{G}_\theta  + \mathbf{k}(\mathbf{s})' \mathbf{M}_\theta \Bigr) \mathbf{y}\,,
\end{equation*}
where \textcolor{PineGreen}{$\mathbf{u}(\mathbf{s}) = (u_1(\mathbf{s}), \dots, u_q(\mathbf{s}))'$}, $\mathbf{k}(\mathbf{s})= \bigl(k(\mathbf{s},\mathbf{s}_1), \ldots, k(\mathbf{s},\mathbf{s}_n) \bigr)'$, and the matrices $\mathbf{G}_\theta$ and $\mathbf{M}_\theta$ can be found as blocks in the inverse matrix,
$$\begin{bmatrix}
\mathbf{K}_\theta & \mathbf{U} \\
\mathbf{U}' & \mathbf{0}
\end{bmatrix}^{-1}=\begin{bmatrix}
\mathbf{M}_\theta & \mathbf{G}_\theta' \\
\mathbf{G}_\theta & \mathbf{L}
\end{bmatrix} \,,$$
where $\mathbf{K}_\theta= \mathbf{K} + \theta \mathbf{I}_n$, $\theta\ge 0$ is a smoothing parameter, and the value of $\mathbf{L}$ is irrelevant here. Further, $\mathbf{G}_\theta$ and $\mathbf{M}_\theta$ take the explicit forms
\begin{equation}\label{eq:A-B}
\mathbf{G}_\theta= \bigl(\mathbf{U}' \mathbf{K}_\theta^{-1}\mathbf{U} \bigr)^{-1}\mathbf{U}' \mathbf{K}_\theta^{-1}
\quad \mbox{and} \quad
\mathbf{M}_\theta=\mathbf{K}_\theta^{-1}-\mathbf{K}_\theta^{-1}\mathbf{U}  \bigl(\mathbf{U}' \mathbf{K}_\theta\mathbf{U}\bigr)^{-1}\mathbf{U}' \mathbf{K}_\theta^{-1}\,,
\end{equation}
which results in $\mathbf{g}=\mathbf{G}_\theta\mathbf{y}$ (i.e., the GLS estimate of the fixed effects in $\mathbf{U}$) and $\mathbf{m}=\mathbf{M}_\theta \mathbf{y}$. 
In the stochastic process setting, $\mathcal{H}_1$ is a space of random variable and not functions as in RKHS. However, this space is isometrically isomorphic to $\mathcal{H}$ under the mapping $ Y(\mathbf{s}) \rightarrow  k(\cdot,\mathbf{s})$ with  $Cov \bigl(Y(\mathbf{s}),Y(\mathbf{s}') \bigr) = k(\mathbf{s},\mathbf{s}')= \left\langle k(\cdot,\mathbf{s}), k(\cdot,\mathbf{s}')\right\rangle $. Hence, there exists a one-to-one inner product correspondence between the two spaces \citep[see][for more details]{Wahba1990}.

\subsection{The reduced-rank random effects model}
After briefly discussing the issues of interpolation and smoothing in the context of RKHS and stochastic processes, the following remarks are crucial for introducing our reduced-rank representation of the conditional model.

\vspace{.3cm}
\noindent \textsl{\textsc{Remark 1}}. The columns of $\mathbf{M}_\theta$ and of $\mathbf{U}$ are orthogonal, i.e. $\mathbf{M}_\theta\mathbf{U}=\mathbf{0}$. This is equivalent to the constraint $\mathbf{U}'\mathbf{m}= \bf 0$ used in Web Appendix A. Also, this implies that the first $q$ eigenvalues of $\mathbf{M}_\theta$ are 0 and the corresponding eigenvectors are given by the $q$ columns of $\mathbf{U}$. 

\vspace{.2cm}
\noindent \textsl{\textsc{Remark 2}}. Let $\theta=0$ and consider the equation $ f_x(\mathbf{s}) =  \Bigl(\mathbf{u}(\mathbf{s})' \mathbf{G}  + \mathbf{k}(\mathbf{s})' \mathbf{M} \Bigr) \mathbf{y} $, 
where $\mathbf{G}$ and $\mathbf{M}$ are as in Equation \eqref{eq:A-B} but with $\mathbf{K}_\theta$ replaced by $\mathbf{K}$. Consider the spectral decomposition of $\mathbf{M}$, $\mathbf{M} =\mathbf{V}{\mathbf{\Lambda}} \mathbf{V}'$, such that $\mathbf{M}\mathbf{v}_l = {\lambda}_l  \mathbf{v}_l$ and $\mathbf{y}=\mathbf{V}\bm{\xi}$. Then, the Kriging predictor can be rewritten as
\begin{equation}\label{eq:low_rank_krig}
f_x(\mathbf{s}) =  \Bigl(\mathbf{u}(\mathbf{s})' \mathbf{G}  + \mathbf{k}(\mathbf{s})' \mathbf{M} \Bigr) \mathbf{V}\bm{\xi} 
= \sum_{l=1}^n \Bigl\{ \bigl(\mathbf{u}(\mathbf{s})' \mathbf{G}  + \mathbf{k}(\mathbf{s})' \mathbf{M} \bigr) \mathbf{v}_l \Bigr\} \xi_l\,.
\end{equation}
Let  $\bm{\psi}_l(\mathbf{s})=\bigl(\mathbf{u}(\mathbf{s})' \mathbf{G}  + \mathbf{k}(\mathbf{s})' \mathbf{M} \bigr) \mathbf{v}_l$ define the $l$-th \textit{principal kriging function} \citep[PKF;][]{kent2001functional, fontanella2019offset}. For  any $\mathbf{s} \in   \lbrace \mathbf{s}_1, \dots, \mathbf{s}_n \rbrace $, this function acts as an interpolator since it satisfies the condition  $\bm{\psi}_l(\mathbf{s})=\mathbf{v}_l(\mathbf{s})$ for $ l=1,\ldots,n$. If the linear combination in Equation \eqref{eq:low_rank_krig} is restricted to a limited  number of eigenvectors, say $p<n$, Equation \eqref{eq:semiparametric-model} represents a reduced-rank random effects model. Furthermore, given that $\mathbf{M}_\theta{\mathbf{U}}=\mathbf{0}$, Equation \eqref{eq:semiparametric-model} can be more explicitly expressed as follows: 
\begin{eqnarray}\label{eq:low_rank_model}
Y(\mathbf{s}_i)&=& \sum_{l=1}^{q}  u_{l}(\mathbf{s}_i) \xi_{l}^a + \sum_{l=q+1}^{p}  \psi_{l}(\mathbf{s}_i) \xi_l^b + \widetilde{\epsilon}_y (\mathbf{s}_i)
\end{eqnarray}
which implies 
$E \left[Y(\mathbf{s}_i) \right]=\sum_{l=1}^q u_{l}(\mathbf{s}_i) \xi_{l}^a,\ 
Cov \bigl(Y(\mathbf{s}_i),Y(\mathbf{s}_j) \bigr)=\bm{\psi}_p(\mathbf{s}_i)' \bm{\Lambda}_p\bm{\psi}_p(\mathbf{s}_j)+\widetilde{\sigma}^2_{\epsilon} I\bigl(\mathbf{s}_i=\mathbf{s}_j \bigr)$, 
where $\bm{\psi}_p(\mathbf{s}_i)=\bigl({\psi}_1(\mathbf{s}_i), \ldots, {\psi}_p(\mathbf{s}_i)\bigr)'$, $\bm{\Lambda}_p=diag\bigl(\lambda_1, \ldots, \lambda_p\bigr)$ and $I\bigl(\mathbf{s}_i=\mathbf{s}_j\bigr)=1$ if $\mathbf{s}_i=\mathbf{s}_j$, and $0$ otherwise. The principal kriging functions can be arranged from coarse scale to fine scale in terms of variation. Hence, by setting the smoothing parameter  equal to zero, model flexibility is controlled by the basis dimension,
so that model selection becomes a matter of choosing $p$ rather than estimating a smoothing parameter $\theta$. It is worth noting that our decomposition, which is based on principal kriging functions, shares similarities with the thin plate regression spline (TPRS) approach described in \citet{wood2003tprs}, who leverages the spectral decomposition of $\mathbf{K}$. Specifically, if $\mathbf{K}\bm{\widetilde{\varphi}}_l = \widetilde{\lambda}_l  \bm{\widetilde{\varphi}}_l$, \textcolor{PineGreen}{with $\widetilde{\lambda}_l$ being the $l$-th largest eigenvalue of $\mathbf{K}$ and $ \bm{\widetilde{\varphi}}_l = \bigl(\widetilde{\varphi}_l(\mathbf{s}_1), \dots, \widetilde{\varphi}_l(\mathbf{s}_n) \bigr)' $ denoting the corresponding eigenvector}, the low-rank representation takes the form:
\begin{eqnarray}\label{eq:TPSwood}
	f_x(\mathbf{s}_i) & = & \sum_{l=1}^{q}  u_{l}(\mathbf{s}_i) \xi_{l}^a + \sum_{l=1}^p \widetilde{\lambda}_l \widetilde{\varphi}_l(\mathbf{s}_i) \xi_l^b\,, \nonumber 
	\end{eqnarray}
subject to the linear constraints $\mathbf{U}' \bm{\widetilde{\varphi}}_l \xi_l^b = \mathbf{0}$, for $l=1,\dots, p$, which is incorporated through a QR decomposition. This is different from our approach. Additionally, while the TPRS bases can be ordered naturally by scale, they are not orthogonal.

\vspace{.2cm}
\noindent \textsl{\textsc{Remark 3}}. Depending on the specific form of the bases $\mathbf{u}(\mathbf{s})$  and the choice of the kernel function  $k(\cdot,\cdot)$, various formulations of the conditional model can be proposed. In this paper, we consider the ``null space'' $\mathcal{H}_0 = span\bigl(1, \mathbf{s}[1], \mathbf{s}[2]\bigr)$, referred to as a \textit{type-1 null space}. Other formulations such as $\mathcal{H}_0 = span\bigl(1, X(\mathbf{s}), \mathbf{s}[1], \mathbf{s}[2]\bigr)$ --- referred to as a \textit{type-2 null space} --- or $\mathcal{H}_0 = span\bigl(1, X(\mathbf{s})\bigr)$ --- referred to as a \textit{type-3 null space} --- \textcolor{PineGreen}{contribute only marginally or are not effective in mitigating the spatial confounding problem}. This occurs because $X$ is contained within the ``null space" $\mathcal{H}_0$, making it difficult for the model to distinguish between the effect of $X$ and the spatial pattern driven by unmeasured spatial factors (see Section \ref{sec:why}). Moreover, given a valid kernel function $k(\cdot,\cdot)$, and due to the orthogonality constraint $\mathbf{M}_\theta\mathbf{U}=\mathbf{0}$, assuming that in model \eqref{eq:low_rank_model} $E \left[Y(\mathbf{s}_i) \right]$ is a polynomial in the exposure, i.e., $\mathbf{u}(\mathbf{s}_i)=\bigl(1, X(\mathbf{s}_i) \bigr)'$, leads to a model structure that is equivalent to the RSR model.

\section{Bias Reduction with Low-Rank Smoothers}\label{sec:why}
To explore under which conditions the confounding bias can be mitigated by the inclusion of basis functions, we compare two models under the ordinary least-squares framework: a non-spatial model without confounding adjustment (hereafter \textit{unadjusted OLS model}), and a model (referred to as \textit{adjusted OLS model}) \textcolor{PineGreen}{with design matrix $\Bigl[	\Tilde{\mathbf{X}} \ \, \mathbf{B} \Bigr]$, where $\mathbf{B}$ is a basis matrix constructed from the PKFs given the kernel function as in Section \ref{sec:model}.}

Following \citet{paciorek2010importance} and \citet{page2017estimation}, let $\widehat{\boldbeta}_{OLS}$ indicate the unadjusted OLS estimator for $\beta_x$, then its bias is equal to 
$$ \Delta_{OLS} = E\left[\widehat{\boldbeta}_{OLS}|\mathbf{X=x}\right] - \boldbeta = \delta \frac{\sigma_w}{\sigma_x} \left( \Tilde{\mathbf{X}}' \Tilde{\mathbf{X}} \right)^{-1} \Tilde{\mathbf{X}}' \mathbf{R}_{\phi_w}^{1/2} \mathbf{R}_{\phi_x}^{-1/2} \mathbf{x}\,, $$
while under the assumptions made in Section \ref{sec:2}, we prove that the bias of the adjusted OLS estimator is $ \Delta_{adj} = \Delta_{OLS} + d_x $, where $d_x$ is the second element of the following vector:
\begin{equation} \label{eq:d_OLS}
	\left[\begin{array}{c} d_0 \\ d_x \end{array}\right] =  \delta \frac{\sigma_w}{\sigma_x} \mathbf{T B'} \biggl\{ \mathbf{\Tilde{X}} (\mathbf{\Tilde{X}'\Tilde{X}})^{-1} \mathbf{\Tilde{X}}' - \mathbf{I}_n \biggr\} \mathbf{R}_{\phi_w}^{1/2} \mathbf{R}_{\phi_x}^{-1/2} \mathbf{x}\,,
\end{equation}
where $\bf B$ is an $n \times (p-1)$ matrix with $\mathbf{b}(\mathbf{s}_i)'$ as its $i$-th row, and $\mathbf{T}$ is a matrix obtained when inverting the cross-product of the design matrix $\Bigl[	\Tilde{\mathbf{X}} \ \, \mathbf{B} \Bigr]$. A complete proof can be found in Web Appendix B. Equation (\ref{eq:d_OLS}) holds for \textit{any} basis expansion, also for that used by \citet{keller2020selecting}.

Bias reduction is achieved when $ |\Delta_{adj}| < |\Delta_{OLS}| $, so it is required that the difference $d_x$ and the bias of the unadjusted OLS estimator must have opposite signs, and that $|d_x| \le |\Delta_{OLS}|$. 

\begin{figure}
	\centering
	\includegraphics[width=14cm]{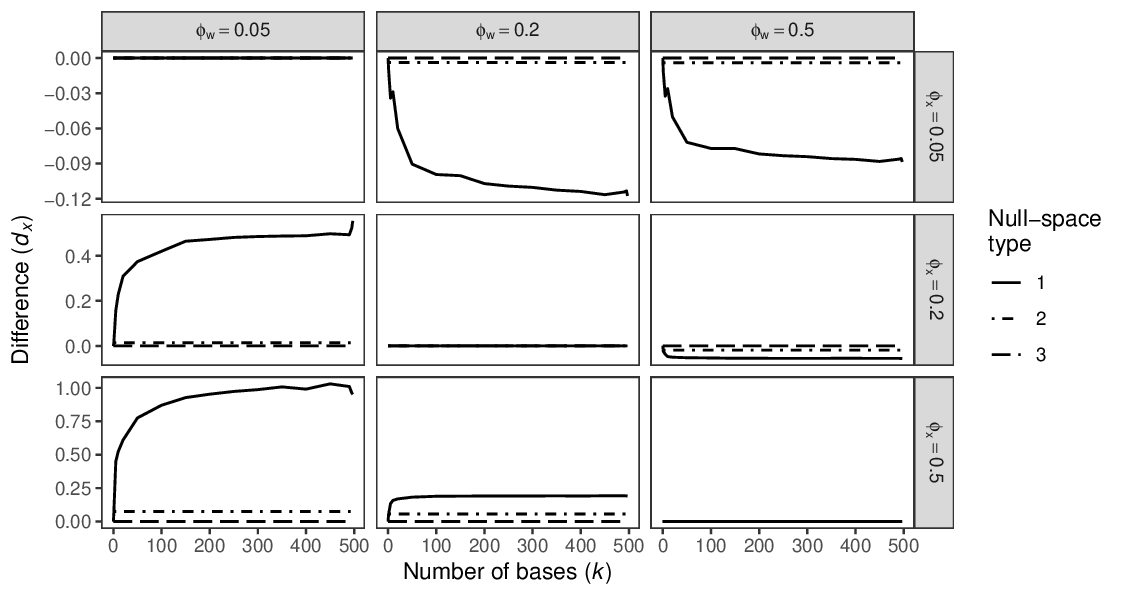}
	\caption[Adjusted vs unadjusted OLS models: amount of bias reduction or amplification]{Adjusted \textit{vs} unadjusted OLS models: amount of bias reduction or amplification, quantified by the difference $d_x$, for increasing number of bases in $\mathbf{B}$ and for several combinations of range parameters. Each line type represents a type of null space used to construct $\mathbf{B}$. The scales on the y-axes are different for each ``row".}
	\label{fig:difference-part1-minus-part2basis-from-1-to-k}
\end{figure}

To understand the relationship between the difference, $d_x$, and the number of bases in $\mathbf{B}$, and how it is affected by the spatial parameters, we run a numerical experiment with the first $k$ principal functions, for $k = 1,\dots, n-3$. Within the unit square, we generate the exposure, $\bf X$, from its marginal distribution while assuming that $n=500$, $\delta=0.5$, $\sigma_w=\sigma_x=1$, and $\phi_w, \phi_x \in \{0.05,0.2,0.5\}$. Hence, the corresponding practical ranges are approximately $0.15$, $0.6$ and $1.5$, respectively. The results are shown in Figure \ref{fig:difference-part1-minus-part2basis-from-1-to-k}. When the bases are constructed from the type-1 null space, spatial confounding is alleviated (i.e., $d_x<0$) only if $ \phi_x < \phi_w $. Besides, as $k$ increases, $d_x$ changes at decreasing rates (however, the model is more likely to overfit the data for large $k$). This implies that the first, low-frequency bases are the most effective in reducing the bias, but they are also the most problematic when $\phi_x > \phi_w$. If $\phi_x = \phi_w$, the confounding bias can not be alleviated as $d_x=0$ for any $k$. 

\textcolor{PineGreen}{Section \ref{sec:3} introduced two more types of null spaces, arguing against their adoption. This can be corroborated also from a simulation standpoint.} As shown in Figure \ref{fig:difference-part1-minus-part2basis-from-1-to-k}, in the case of type-2 null space, the confounding bias can be alleviated (or amplified) \textit{only} through the inclusion of the spatial coordinates in the model, whereas in the case of the third type, mitigating confounding issues is not possible (see Web Appendix B for a proof).

In summary, adopting the type-1 null space seems to provide the best results in terms of confounding bias mitigation, since the exposure and the bases are not orthogonal, making it possible to capture the correlation between $\mathbf{X}$ and the unmeasured spatial factor. 
However, we seek for a parsimonious model that allows for the selection of the most important bases, therefore in the next Section we propose a regularization strategy.

\section{Proposed Regularization of the Regression Model}
\label{sec:model}

{\color{PineGreen}\makeatletter\let\default@color\current@color\makeatother
Basing our analysis on the type-1 null space, we reformulate Equation \eqref{eq:semiparametric-model} as a \textit{partial thin plate spline model} \citep[see, e.g.,][]{Wahba1990}:
\begin{eqnarray}
Y(\mathbf{s}_i)&=& \beta_0+\beta_x X(\mathbf{s}_i)+ \widetilde{f}(\mathbf{s}_i) + \widetilde{\epsilon}_y (\mathbf{s}_i) \nonumber \\
&=& \beta_0 + \beta_x X(\mathbf{s}_i)+ \sum_{l=2}^{q}  u_{l}(\mathbf{s}_i) \xi_{l}^a + \sum_{l=q+1}^{p}  \psi_{l}(\mathbf{s}_i) \xi_l^b + \widetilde{\epsilon}_y (\mathbf{s}_i) \nonumber \\
&=& \beta_0 + \beta_x X(\mathbf{s}_i) + \mathbf{b}(\mathbf{s}_i)' \widetilde{\bm{\xi}} + \widetilde{\epsilon}_y (\mathbf{s}_i)\,,\label{eq:low_rank_model2}
\end{eqnarray}
where the basis functions $ u_j(\mathbf{s}_i)$ are monomials in the spatial coordinates of $\bf s$, such that $q=3$ and $\mathbf{u}(\mathbf{s}_i)=\bigl(1, \mathbf{s}_i[1], \mathbf{s}_i[2] \bigr)'$, $\mathbf{b}(\mathbf{s}_i) = \bigl( u_{2}(\mathbf{s}_i), u_{3}(\mathbf{s}_i), \psi_{q+1}(\mathbf{s}_i), \dots, \psi_{p}(\mathbf{s}_i) \bigr)'$, and $\widetilde{\bm{\xi}} = \bigl(\xi_2^a, \dots, \xi_q^a, \xi_{q+1}^b, \dots, \xi_p^b\bigr) '$ is a $(p-1)$-dimensional vector of expansion coefficients. Next, we define the kernel function as $k(\mathbf{s}_i, \mathbf{s}_j)=\frac{1}{8\pi} \| \mathbf{s}_i - \mathbf{s}_j \|^2 \log \| \mathbf{s}_i - \mathbf{s}_j \|$. By assuming a correspondence between the reproducing kernel $k(\mathbf{s}, \mathbf{s}')$ and $Cov \bigl(Y(\mathbf{s}),Y(\mathbf{s}') \bigr)$, $\forall \mathbf{s}, \mathbf{s}' \in \mathcal{S}$ (see Section \ref{subsec:smoothing}), the link between thin plate splines and kriging \citep[see][for more formal details]{Kent1994} shows that $\widetilde{f}(\mathbf{s}_i)$ defines a \textit{thin plate spline}. 

The exposure $X(\mathbf{s}_i)$ is not included in the ``null space" $\mathcal{H}_0$. This is a key feature of the proposed approach, crucial to avoid the problematic implications due to the orthogonality constraint --- see Remark 3. A second important feature is that the bases from the thin-plate spline act as proxy variables for the spatially-varying unmeasured confounder.
}


The likelihood function is obtained through the model in Equation \eqref{eq:low_rank_model2}.
As inference is performed following the Bayesian paradigm, a key aspect of model specification is the prior distribution for the coefficients of the basis functions. We propose using spike-and-slab priors \citep{george1997approaches}, extensively adopted to identify the most important covariates associated with the outcome, and to estimate posterior model probabilities \citep{fontanella2019offset}. These priors offer a flexible alternative to truncation \citep{morris2015functional}, allowing to explore all possible models within a hierarchical formulation, at a minimal computational cost. 

In this paper, we do not truncate the basis matrix (so $p=n-1$) and consider a \textit{non-local} spike-and-slab prior structure known as the \textit{first-order product moment} (\textit{pMOM}) priors \citep{johnson2012bayesian, rossell2017nonlocal}. \textit{Local} prior structures are characterized by a prior density \textcolor{PineGreen}{which does not decay to zero in a neighborhood of the origin. Conversely, \textit{non-local} prior structures assign all the prior probability mass outside a neighborhood of zero for any slab component. This property permits model selection procedures based on these priors to eliminate regression models that contain unnecessary bases}. In high-dimensional regression settings, classes of non-local priors are proposed by \citet{johnson2012bayesian} and \citet{rossell2017nonlocal}, with the main advantage of not demanding a high computational cost. 
Let $M_t$, $t=1,\dots,2^{n-1}$, denote the $t$-th member of the model space given by all possible combinations of the $n-1$ bases. According to the pMOM priors, the $l$-th element of $\widetilde{\bm{\xi}}$ is considered as a mixture between a point mass at zero and a bimodal density, namely
$
	\xi_l|\widetilde{\sigma}_\epsilon^2,M_t \stackrel{ind}{\sim} \xi_l^2(\nu \widetilde{\sigma}_\epsilon^2)^{-1} N\left( 0, \nu \widetilde{\sigma}_\epsilon^2 \right) 
$, $l=1,\dots,n-1$, where $\nu$ is a scale parameter. Additionally, a Beta-Binomial$(1,1)$ prior is assumed on $M_t$. 

We assign independent priors on the remaining parameters, namely $\beta_0, \beta_x \stackrel{iid}{\sim} N(0, V_\beta)$, and $\widetilde{\sigma}_\epsilon^2 \sim IG(a, b)$, where $V_\beta$, $a$ and $b$ are known. Hence, the parameter vector with all the unknowns, $\bm{\Theta}$, contains $\boldbeta$, $\widetilde{\boldxi}$, and $\widetilde{\sigma}_\epsilon^2$. The posterior distribution of $\bm{\Theta}$ is proportional to the likelihood times the prior distribution. The Markov chain Monte Carlo (MCMC) algorithm described in \citet{rossell2017nonlocal} is used to get posterior samples, and is available through the R package \verb|mombf| \citep{R, mombf}. 

In Web Appendix C, we also consider two variants of local prior structures, based on the works by \citet{george1997approaches} and \citet{ishwaran2005spike}. However, the proposal imposing pMOM priors outperform both variants (see Web Appendix D). 


\section{Simulation Study} \label{sec:simulations}
We conduct an extensive simulation study to shed light on the ability of the proposed approach (abbreviated as \textit{SS\_mom}) to mitigate the spatial confounding bias, and to compare it to other methods in the literature. The data, sampled from a $ 64 \times 64 $ grid over the unit square, are generated by fixing $\delta=0.5$ and $\sigma_x^2=1$, and by considering a grid of values for the range parameters, $ \phi_x, \phi_w  \in [0.05, 0.5] \times [0.05, 0.5]$ with steps of $0.05$. Each combination of these values is referred to as a \textit{configuration}.
In each configuration, the exposure is generated only once, then conditional on it, we sample $ \mathbf{W}_x $ and the outcome, $\mathbf{Y}$, to obtain 100 different \textit{replicates}. We set $ \boldbeta =(1,2)' $ and $ \sigmaeps^2=0.25 $. To make comparisons across configurations, we consider $ n=500 $ randomly-sampled locations as fixed, and we set the relative OLS bias (i.e. the ratio $\Delta_{OLS}/\beta_x$) to $0.15$. Therefore, given $ \delta $, we allow $\sigma_w^2$ to vary accordingly. The proposed approach is compared to several existing methods, which are reported in Table \ref{tab:competing-approaches}.

\begin{table}
	\caption{Competing approaches analyzed in the simulation study.} \label{tab:competing-approaches}
	\centering
	\renewcommand{\arraystretch}{1.5}
	\begin{tabular*}{\columnwidth}{p{2.5cm} p{9cm} p{3.8cm}}
		\hline
		\textbf{Method} & \textbf{Description} & \textbf{Reference} \\
		\hline
		OLS & The unadjusted non-spatial model, fitted using ordinary least squares. & \\
		SRE (spatial random effect) & The spatial linear model in Equation (\ref{eq:main}) where $W(\mathbf{s}_i)$ is an Gaussian process with exponential covariance function, independent of the exposure. & \\
		SpatialTP & The model where $W(\mathbf{s}_i)$ is approximated by a basis expansion from a thin-plate regression spline (TPRS) evaluated at the data locations. & \citet{dupont2022spatialplus}\\
		gSEM (geoadditive Structural Equation Model) & The model where the spatial dependence is regressed away from both the response and the exposure through the use of TPRS. Then, a regression involving the residuals is used to estimate the exposure effect. & \citet{thaden2018structural},\newline \citet{dupont2022spatialplus} \\
		Spatial+ & The model where the spatial dependence is regressed away only from the exposure. Then, the outcome is regressed on the residuals and on a smooth term. A penalty parameter is estimated using generalized cross-validation. & \citet{dupont2022spatialplus} \\
		Spatial+\_fx & The Spatial+ model where the penalty parameter is fixed at zero. & \citet{dupont2022spatialplus} \\
		KS (Keller-Szpiro) & This approach consists in: (i) regressing the outcome on a matrix of TPRS basis functions, with its degrees of freedom found by minimizing the Akaike information criterion (AIC); (ii) fitting a SpatialTP model where the TPRS has exactly the pre-selected degrees of freedom. & \citet{keller2020selecting} \\
		SA (spectral adjustment) & The model where a basis expansion of B-splines constructed in the spectral domain and a Gaussian process, independent of the other covariates, are included in the model specification. & \citet{guan2023spectral} \\
		\hline
	\end{tabular*}
\end{table}

To avoid scale issues, all variables are standardized, but for ease of interpretation, results are shown on the original scale of the coefficients. The following values are adopted for the hyperparameters in the model: $ V_\beta = 1 $, $ a=2 $, $b=0.1 $, and $ \nu = 0.348 $ as suggested by \citet{johnson2012bayesian}.

\subsection{Results}
\label{subsec:sim_results}

\begin{figure}
	\centering
	\includegraphics[width=\textwidth]{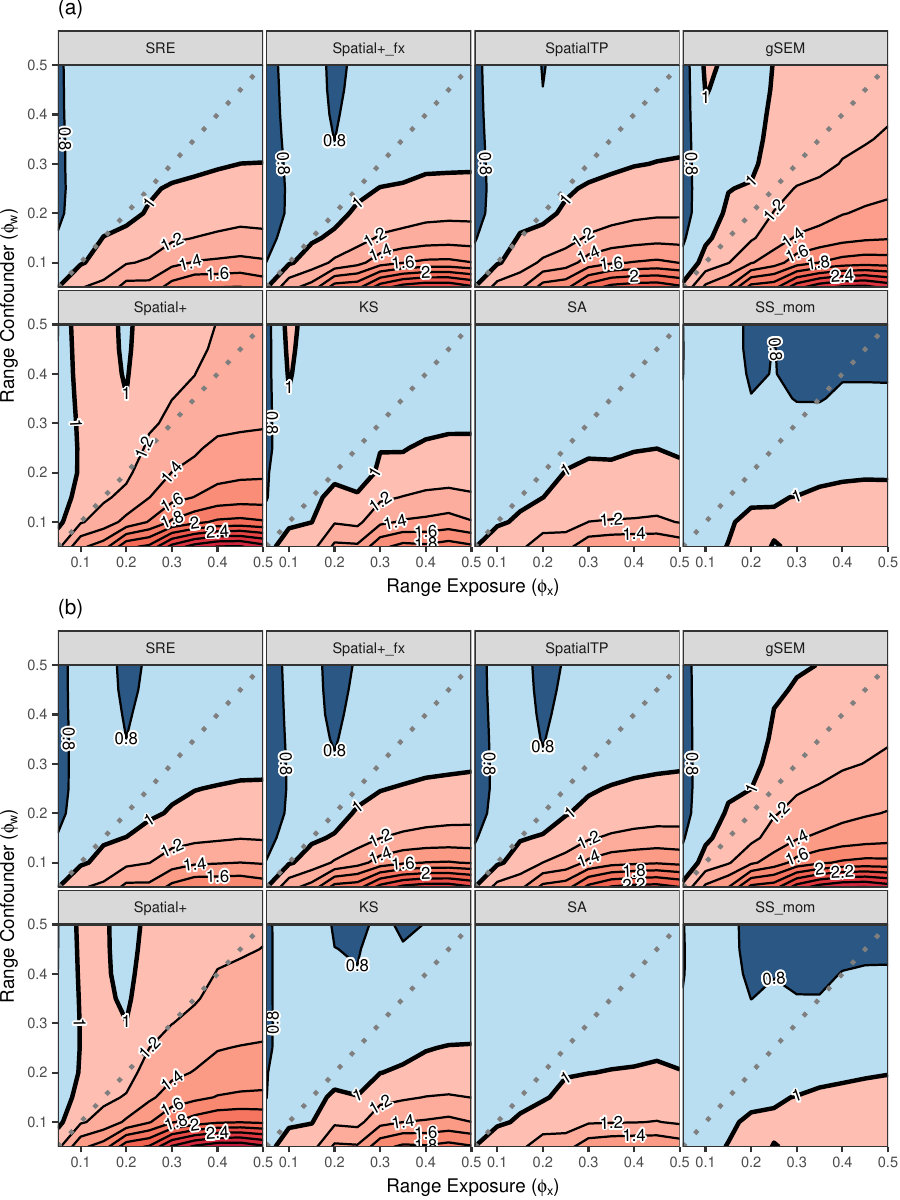}
	\caption[Contour plots]{Contour plots showing ratios between competing methods and the OLS model, in terms of MAE (a) and RMSE (b). The thicker lines represent the value 1. Values smaller than 1 are colored in shades of blue, while values greater than 1 are colored in shades of red. Darker colors indicate values further from 1. The gray dotted line represents the diagonal line $ \phi_x = \phi_w $.}
	\label{fig:contours}
\end{figure}

We adopt OLS as the reference model, and all other methods are compared to it by computing two ratios, which we shall denote as $Q_1$ and $Q_2$. They are related to the \textcolor{PineGreen}{mean absolute error (MAE)} and the root mean squared error (RMSE) of the estimator for $\beta_x$, respectively. The reference model is always at the denominator, hence, under the simulation setup, ratios smaller than 1 indicate that the method under consideration performs better than OLS, and vice versa. In Figure \ref{fig:contours} the areas colored in shades of blue (red) indicate ratios less (greater) than 1. Darker colors indicate a more pronounced benefit (or disadvantage). We expect for the best method to have the blue area maximized, i.e. the ratio should be smaller than 1 in the majority of configurations. 

The ratio $Q_1$ is obtained, \textcolor{PineGreen}{as the ratio of MAE of each method over that of the OLS model. The MAEs are calculated over the $100$ replicates}. Figure \ref{fig:contours}a thus shows the ratios for each $ \phi_x, \phi_w $ pair and for each method. A ratio below 1 indicates that the competing approach reduces the confounding bias with respect to OLS, while a ratio above 1 is an indication of bias amplification. A pattern common to all panels is that ratios smaller than 1 are more frequent when $ \phi_x< \phi_w $. However, the SS\_mom model is the only one that is able to reduce the bias also when $ \phi_x> \phi_w $ and $ \phi_x< 0.15 $. Furthermore, the thicker contour line (corresponding to $Q_1=1$) only partially overlaps the diagonal line $ \phi_x = \phi_w $. That is, if the range parameters are equal and greater than 0.3, the bias can be mitigated by using all approaches except gSEM and Spatial+, which report bias amplification.

\textcolor{PineGreen}{To compare the methods quantitatively, we consider the probability $\Pr\{abs(\widehat{\beta}_x^* - \beta_x)<abs(\widehat{\beta}_{x,OLS} - \beta_x)\}$, where $\widehat{\beta}_x^*$ is the estimate of $\beta_x$ from any of the competing methods, $\widehat{\beta}_{x,OLS}$ is the OLS estimate, and $abs(\cdot)$ denotes the absolute value function. The probability is estimated by averaging the proportions of occurrences in which the inequality holds across all Monte Carlo replicates and settings. Table \ref{tab:ratios} shows that the SS\_mom model has the highest probability of reducing the bias ($72\%$), followed by the SA model ($52\%$). In most practical situations, the unmeasured confounder is spatially smoother than the exposure. Moreover, we may want to exclude very rough spatial variations in the exposure. Hence, the probabilities $\Pr\{abs(\widehat{\beta}_x^* - \beta_x)<abs(\widehat{\beta}_{x,OLS} - \beta_x) | \phi_x < \phi_w\}$ and $\Pr\{abs(\widehat{\beta}_x^* - \beta_x)<abs(\widehat{\beta}_{x,OLS} - \beta_x) |0.2 < \phi_x < \phi_w\}$ are also reported in Table \ref{tab:ratios}. The proposed model achieves again the highest probability, $74\%$ in both cases.}


The comparison based solely on the bias does not account for the variability of the estimates. Therefore, $Q_2$ is defined as the ratio between the RMSE of $\beta_x$ of any competing approach and the RMSE of the OLS model.
The results are shown in Figure \ref{fig:contours}b --- which qualitatively confirms the patterns found in Figure \ref{fig:contours}a --- and in Table \ref{tab:ratios}. The SS\_mom model has the highest probability of reducing the RMSE ($81\%$). To consider conditions of more pronounced RMSE reduction or amplification, other performance metrics are computed. They comprise the probability of an RMSE reduction greater than $20\%$, which is the highest for the SS\_mom model when $ \phi_x < \phi_w $ ($42\%$), and when $ 0.2 < \phi_x < \phi_w $ ($80\%$). The last metric is the probability of a pronounced RMSE amplification (greater than $80\%$), which is the lowest for SS\_mom, SA, and SRE models ($0\%$).


\begin{sidewaystable}
	\centering
	\caption{Performance metrics for the competing approaches and for both ratios analyzed, $Q_1$ and $Q_2$. The best performing method is highlighted in italics.}
	\label{tab:ratios}
	\begin{tabular}{lllllllll}
	\hline
				& SRE        & Spatial+\_fx & SpatialTP & gSEM & Spatial+ & KS   & SA         & \textbf{SS\_mom}  \\ \hline
	$\Pr\{abs(\widehat{\beta}_x^* - \beta_x)<abs(\widehat{\beta}_{x,OLS} - \beta_x)\}$   %
				& 0.48       & 0.51         & 0.48      & 0.40 & 0.32     & 0.51 & 0.52       & \textit{0.72} \\
	$\Pr\{abs(\widehat{\beta}_x^* - \beta_x)<abs(\widehat{\beta}_{x,OLS} - \beta_x)|\phi_x < \phi_w\}$ %
		      	& 0.66       & 0.68         & 0.66      & 0.58 & 0.49     & 0.66 & 0.63          & \textit{0.74} \\
	$\Pr\{abs(\widehat{\beta}_x^* - \beta_x)<abs(\widehat{\beta}_{x,OLS} - \beta_x)|0.2 < \phi_x < \phi_w \}$ %
				& 0.58       & 0.62         & 0.57      & 0.49 & 0.40     & 0.60 & 0.61          & \textit{0.74} \\
	$\Pr(Q_2<1)$                          & 0.68       & 0.66         & 0.67      & 0.32 & 0.13     & 0.71 & 0.73       & \textit{0.81} \\
	$\Pr(Q_2<0.8|\phi_x < \phi_w)$        & 0.22       & 0.27         & 0.27      & 0.18 & 0.04     & 0.27 & 0.00       & \textit{0.42} \\
	$\Pr(Q_2<0.8|0.2 < \phi_x < \phi_w)$  & 0          & 0            & 0         & 0    & 0        & 0.2  & 0          & \textit{0.8}  \\
	$\Pr(Q_2>1.8)$                        & \textit{0} & 0.07         & 0.05      & 0.1  & 0.11     & 0.04 & \textit{0} & \textit{0}    \\ \hline
	\end{tabular}
\end{sidewaystable}

Web Appendix D gives more insights for some selected configurations, showing how the approaches differ in terms of effect estimates \textcolor{PineGreen}{(see Figure 1 in the Supplementary Materials)} and number of bases selected. We also compare the SS\_mom model to other methods based on local prior structures, and show that the difference $d_x$ (given the selected bases) is always smaller or equal to zero under pMOM priors. Furthermore, Web Appendix C reports other simulation scenarios, where we also vary other parameters such as $\delta$, $\sigma_x^2$, $\sigmaeps^2$. Finally, the sensitivity of the proposed model to the choice of keeping the exposure fixed across replicates is investigated.

\section{Reducing Confounding Bias in Ozone--NO\textsubscript{x} Association}
\label{sec:application}
We apply the proposed model to an environmental dataset concerning tropospheric ozone (O\textsubscript{3}). This greenhouse gas poses risks to both human health and the environment when present in the troposphere. For example, it may cause respiratory diseases, or permanently damage lung tissue. Additionally, it has the potential to harm agricultural crops and forests \citep{nguyen2022ozone}. This gas is not directly emitted, but rather forms through chemical reactions between nitrogen oxides (NO\textsubscript{x}) and volatile organic compounds (VOCs), in the presence of sunlight and high temperatures. Consequently, ozone concentrations typically peak during daytime in summer \citep{nguyen2022ozone,nussbaumer2022ozone}.

The relationship between O\textsubscript{3} and nitrogen oxides is inferred by considering their average values during summer 2019 (from June to August) in three Italian regions, namely Lazio, Abruzzo and Molise. NO\textsubscript{x} is made up of nitrogen oxide (NO) and nitrogen dioxide (NO\textsubscript{2}), and is released into the atmosphere from combustion sources (e.g., industrial facilities and motor vehicles). To control for potential confounders, we employ the following model, for $i=1, \dots, n$ \citep{chen2020meteorological,nguyen2022ozone,silva2022ozone}:
\begin{align}
	log(O_3 (\mathbf{s}_i)) &= \beta_1 log(NO_x (\mathbf{s}_i)) + \beta_2 u_{10} (\mathbf{s}_i) + \beta_3 v_{10} (\mathbf{s}_i) + \beta_4 Temp(\mathbf{s}_i) + \beta_5 SSR(\mathbf{s}_i) \nonumber \\ 
	& + \beta_6 VOC(\mathbf{s}_i) + \beta_7 RH(\mathbf{s}_i) + \epsilon(\mathbf{s}_i)\,,\qquad \epsilon(\mathbf{s}_i) \stackrel{iid}{\sim} N(0, \sigma_{\epsilon}^2)\,, \label{eq:ozone}
\end{align}
where $log(O_3 (\mathbf{s}_i))$ is the logarithm of ozone concentrations (in $\mu g/m^3$). The covariates are the logarithm of NO\textsubscript{x} concentrations (in $\mu g/m^3$), eastward, $u_{10}(\cdot)$, and northward, $v_{10}(\cdot)$ components of the wind at 10 meters above Earth's surface (in $m/s$), air temperature at 2 meters above Earth's surface ($Temp(\cdot)$, in $K$), surface net solar radiation ($SSR(\cdot)$, in $J/m^2$), volatile organic compounds ($VOC(\cdot)$, in $\mu g/m^3$), and relative humidity ($RH(\cdot)$, in $\%$).

We collected hourly remote-sensed measurements of NO and NO\textsubscript{2} (which are added to derive NO\textsubscript{x} concentrations), VOC, and O\textsubscript{3} concentrations, on a 0.1° regular latitude-longitude grid, from the Copernicus Atmosphere Monitoring Service (CAMS) Atmosphere Data Store \citep{CAMSdataset}. We considered only the time span between 8 a.m. and 6 p.m., while the grid points are $n=353$. A similar procedure is followed to gather meteorological variables, with data available at the same spatial resolution from the ERA5-Land reanalysis dataset within the Copernicus Climate Change Service (C3S) Climate Data Store \citep{munoz2019era5}. Because RH is not directly available, it was derived from dew-point temperature, $DT$, using the August-Roche-Magnus approximation formula \citep{alduchov1996improved}: $	RH = 100 \times \exp \left(\frac{17.625 \times DT}{243.04+DT}-\frac{17.625 \times Temp}{243.04+Temp}\right)$. We use the latitude-longitude coordinates since the study domain is sufficiently small. The satellite data used are already post-processed, so that there are no missing values.

By fitting the ``full" model in Equation (\ref{eq:ozone}) using the OLS method, we obtain that a $1\%$ increase in NO\textsubscript{x} concentration is associated with a $0.021\%$ increase in O\textsubscript{3} on average (95\% confidence interval: $[0.005\%, 0.038\%]$).
Similar estimates are provided by SRE and SS\_mom models, respectively equal to $0.021\%$ (95\% credible interval: $[0.011\%, 0.031\%]$) and $0.019\%$ (95\% credible interval: $[0.012\%, 0.025\%]$). 

\begin{figure}
	\centering
	\includegraphics[width=14cm]{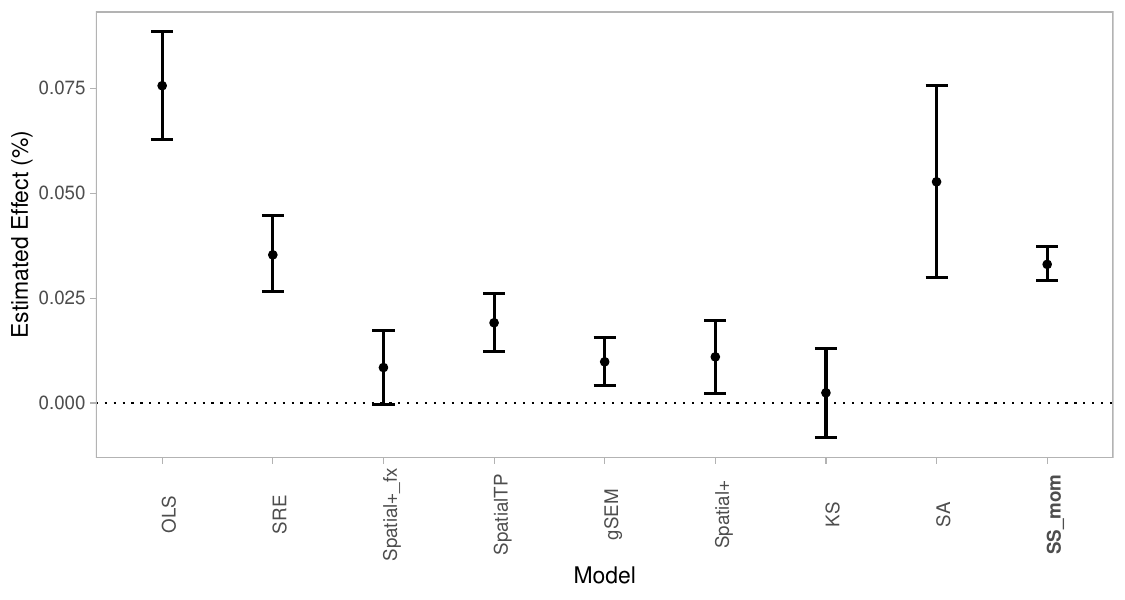}
	\caption[Estimated NO\textsubscript{x}-O\textsubscript{3} association]{Estimated effects (and respective 95\% intervals) of NO\textsubscript{x} on O\textsubscript{3}, under different models. ``OLS" refers to the model where all potential confounders are excluded.}
	\label{fig:o3_estimates}
\end{figure}

We now consider the exposure as the only covariate, with the purpose to explore if the estimates from the competing approaches differ from the ``full'' model's estimate.
The results obtained by all approaches previously compared are depicted in Figure \ref{fig:o3_estimates}. The estimate for the non-spatial model, labeled ``OLS" in the figure, has increased to $0.076\%$ (95\% confidence interval: $[0.063\%, 0.088\%]$). On the other hand, the estimates from SS\_mom and the other competing models are closer to those from the ``full" model. Also, the credible interval of the proposed approach is bounded away from zero. Hence, these results might indicate that spatial confounding bias can be mitigated in this application.


\section{Discussion}
\label{sec:discussion}
In this paper, a novel approach to mitigate spatial confounding has been proposed. We have exploited the principal spline representation of the latent spatial structure, where the basis functions are constructed based on the spectral decomposition of the partial information matrix $\bf M$ \citep{mardia1998kriged}. Within a reduced-rank representation the model in Equation \eqref{eq:semiparametric-model}, we have demonstrated the relationship between the bias from the adjusted OLS model and that from the non-spatial OLS model, which holds for any basis expansion. Reflecting the findings in \citet{paciorek2010importance} and assuming all spatial parameters as known, we show that spatial confounding can be mitigated only when the spatial range of the exposure ($\phi_x$) is smaller than that of the unmeasured confounder ($\phi_w$). However, if the bases are orthogonal to the exposure, $d_x$ in Equation (\ref{eq:d_OLS}) is null (i.e., the bias equals that of the non-spatial model), regardless of the number of bases included.

The proposed approach involves imposing independent spike-and-slab priors on the expansion coefficients. This approach is inherently flexible and allows for the selection of the best proxies from the given set of PKFs. Theoretically, \citet{MalsinerWalli2016comparing} show that when independent spike-and-slab priors are used, the posterior inclusion probabilities tend to increase if the bases are not orthogonal. In this sense, pMOM priors coupled with the principal spline bases have an advantage over TPRS bases. In fact, the SS\_mom searches the region of the model space with the highest posterior probability, which, from simulations, seems to be sparser when the PKFs are used. Hence, as evidenced by \citet{rossell2017nonlocal}, the MCMC sampler becomes very efficient.

We conducted an extensive simulation study, in which SS\_mom outperforms competing approaches (see Section \ref{sec:simulations}). The areas where the ratios $Q_1$ and $Q_2$ are smaller than 1 are maximized, hence the proposed model has the highest probability of reducing the bias and RMSE, and the lowest probability of bias amplification. Besides, it is the only method among the investigated ones able to reduce bias also when $\phi_w < \phi_x < 0.15$.

A common characteristic in Figure \ref{fig:contours} is that the thicker contour lines overlap only partly the bisector of each panel. This may contradict the conclusions of Section \ref{sec:why}, which would suggest a perfect overlap. However, in Section \ref{sec:why} we assumed that all spatial parameters are known, but in practice they are not.

Approaches such as gSEM, Spatial+, and KS require at least two steps to estimate the exposure effect. However, the propagation of uncertainty in parameter estimation from one step to another should be taken into account \citep{schmidt2022discussion}. In contrast, the proposed method is a one-step approach and thus bypasses this issue. The proposed approach shares some similarities with the SA model, in that both models look at spatial confounding from a ``spatial scale'' perspective. Scale selection is not performed \textit{a priori} in either case, but rather through data-driven frameworks. The SA model, however, requires the exposure to be observed or interpolated on a grid \citep{guan2023spectral}, whereas our approach can deal more straightforwardly with data from irregularly-spaced locations.

When analyzing the association between NO\textsubscript{x} and O\textsubscript{3}, we have seen that not accounting for measured confounders leads to an OLS estimate that is $3.6$ times larger than that from the ``full'' model. However, we are aware that the ``full'' model may not recover the true exposure--outcome association. The application was mainly intended to show that the proposed approach can successfully mitigate bias in real-world data settings.

Finally, if the goal is to alleviate the impact of spatial confounding on the exposure--outcome association, we suggest using the proposed low-rank model, where the bases are correlated with the exposure and product moment priors are imposed on the expansion coefficients. In most practical scenarios, $\phi_w$ is grater than $\phi_x$, hence the proposed model is a valuable tool for confounding bias mitigation.

\section*{Acknowledgements}

The authors are grateful to the two reviewers and Associate Editor and Co-Editor for their constructive comments. \citet{munoz2019era5} was downloaded from the \citet{copernicus2023era5}. \citet{CAMSdataset} was downloaded from the \citet{copernicus2023cams}. The results contain modified C3S and CAMS information. Neither the European Commission nor ECMWF is responsible for any use that may be made of the Copernicus information or data it contains. We acknowledge \citet{dupont2022spatialplus} and \citet{guan2023spectral} for making their code available. The research of C.Z., P.V. and L.I. is funded by the European Union - NextGenerationEU. C.Z.: research project PRIN PNRR 2022 (Bando 2023) \textit{SLIDE - Stochastic Modeling of Compound Events} (P2022KZJTZ - CUP D53D23018920001). P.V. and L.I.: research project PRIN2022 \textit{CoEnv - Complex Environmental Data and Modeling} (2022E3RY23 - CUP D53D23011080006). A.M.S. is grateful for financial support from the Natural Sciences and Engineering Research Council (NSERC) of Canada (Discovery Grant RGPIN-2017-04999).\vspace*{-8pt}


\section*{Supplementary Materials}

The R code to reproduce the analyses is available at \url{https://github.com/czaccard/spatial_confounding_SS.git}.\vspace*{-8pt}



\bibliographystyle{abbrvnat} 
\bibliography{references}

\appendix

\section{Smoothing and interpolation in an RKHS}\label{sec:RKHS}

In this section, we look at the optimal smoothing problem in the functional setting. Let $\mathcal{H}$ be an RKHS with \textit{reproducing kernel} (RK) $k(\cdot,\cdot)$ and suppose that $\mathcal{H}=\mathcal{H}_0 \bigoplus \mathcal{H}_1$, where $dim(\mathcal{H}_0) = q$, say, is finite. 
Let $ \lbrace u_j(\mathbf{s}_i), j = 1,\dots, q \rbrace $ form a basis of $\mathcal{H}_0$. Finally, let $\mathcal{P}_1$ denote orthogonal projection onto $\mathcal{H}_1$. Then, it is well known (\citealp{Wahba1990}; pp. 10-14) that the function $f_x(\mathbf{s})$ minimizing
$$\sum_{i=1}^n \Bigl(y(\mathbf{s}_i) -  f_x(\mathbf{s}_i)\Bigr)^2 + \theta \bigl( \mathcal{P}_1 \bm{f}\bigr), $$
where $\theta$ is a smoothing parameter, takes the form
\begin{eqnarray*}\label{eq:f_rkhs}
f_x(\mathbf{s}) &=& \sum_{l=1}^q g_l u_l(\mathbf{s}) + \sum_{j=1}^n m_j k(\mathbf{s},\mathbf{s}_j) \\
&=& \bm{u}(\mathbf{s})' \bm{g} + \bm{k}(\mathbf{s})' \bm{m}\,,
\end{eqnarray*}
with $\bm{k}(\mathbf{s})= \bigl(k(\mathbf{s},\mathbf{s}_1), \ldots, k(\mathbf{s},\mathbf{s}_n) \bigr)'$. Here, the vectors of coefficients $\bm{g}$ and $\bm{m}$ satisfy the equations
$$\bm{U} \bm{g} + \bm{K}_\theta \bm{m} = \bm{y}, \quad \quad \bm{U}'\bm{m}=\bm{0}$$
where $\bm{K}_\theta= \bm{K} + \theta \bm{I}$, with $\bm{K}$  and $\bm{U}$ being $(n \times n)$ and  $(n \times q)$ matrices with elements $k_{ij}=k(\mathbf{s}_i,\mathbf{s}_j)$ and $u_{il}=u_{l}(\mathbf{s}_i), \ i,j=1, \ldots, n, \  \ l=1, \ldots, q$. The solution for the coefficients in the smoothing problem are thus
\begin{equation}\label{eq:a}
\bm{g}= \bigl(\bm{U}' \bm{K}_\theta^{-1}\bm{U} \bigr)^{-1}\bm{U}' \bm{K}_\theta^{-1}\mathbf{y}
\end{equation}
and 
\begin{equation}\label{eq:b}
 \bm{m}=\bigl(\bm{K}_\theta^{-1}-\bm{K}_\theta^{-1}\bm{U}  \bigl(\bm{U}' \bm{K}_\theta\widetilde{\bm{U}}\bigr)^{-1}\bm{U}' \bm{K}_\theta^{-1}\bigr) \mathbf{y}.
\end{equation}
Note that equation \eqref{eq:b} yields  $  \bm{m} = \bm{K}_\theta^{-1}\bigl(\bm{y}-\bm{U} \bm{g} \bigr)$, and that as $\theta \rightarrow 0$, $\bm{K}_\theta \rightarrow \bm{K} $, reducing  the smoothing problem to an interpolation problem.

\section{Derivation of Difference Between Unadjusted and Adjusted OLS Estimators} \label{sec:supp-difference-d}
Consider a more compact form for the proposed semi-parametric model:
\begin{equation} \label{eq:semipar_compact2}
	\mathbf{Y} = \Tilde{\mathbf{X}} \boldbeta + \mathbf{B} \widetilde{\boldxi} + \widetilde{\boldepsilon} = \mathbf{A} \boldtheta + \widetilde{\boldepsilon} \,,
\end{equation}
where $\mathbf{A} = \begin{bmatrix} \Tilde{\mathbf{X}} & \mathbf{B} \end{bmatrix} $ indicates the $n \times (p+2)$ full design matrix where $\widetilde{\mathbf{X}}$ and $\mathbf{B}$ are bound column-wise, and  $\boldtheta = (\boldbeta', \widetilde{\boldxi}')'$ is the associated vector of coefficients.

The expected value of the adjusted OLS estimator is:
\begin{align*}
	E\left[ \widehat{\boldtheta}_{OLS}|\mathbf{X=x} \right] &=  (\mathbf{A'A})^{-1} \mathbf{A}' E\left[ \mathbf{Y}|\mathbf{X=x} \right]\\
	&= (\mathbf{A'A})^{-1} \mathbf{A}' \left( \Tilde{\mathbf{X}} \boldbeta + \boldmu_{w|x} \right) \\
	&= (\mathbf{A'A})^{-1} \mathbf{A}' \Tilde{\mathbf{X}} \boldbeta + \delta \frac{\sigma_w}{\sigma_z} (\mathbf{A'A})^{-1} \mathbf{A}' \mathbf{R}_{\phi_w}^{1/2} \mathbf{R}_{\phi_x}^{-1/2} \mathbf{x}
\end{align*}

Since $\mathbf{A'A}$ is a block matrix, it follows that:
\begin{equation} \label{eq:blkmat1}
	(\mathbf{A'A})^{-1} = \left[ \begin{array}{cc}
		\Tilde{\mathbf{X}}' \Tilde{\mathbf{X}} & \Tilde{\mathbf{X}}' \mathbf{B}\\
		\mathbf{B}' \Tilde{\mathbf{X}} & \mathbf{B}'\mathbf{B}\\
	\end{array} \right]^{-1} = \left[ \begin{array}{cc}
		\mathbf{Q} & -\mathbf{T} \\
		\boldXi & \mathbf{S}^{-1}\\
	\end{array} \right]
\end{equation}
and
\begin{equation} \label{eq:E-theta}
	E\left[ \widehat{\boldtheta}_{OLS}|\mathbf{X=x} \right] = (\mathbf{A'A})^{-1} \mathbf{A}' \Tilde{\mathbf{X}} \boldbeta + \delta \frac{\sigma_w}{\sigma_z} \left[ \begin{array}{cc}
		\mathbf{Q} & -\mathbf{T} \\
		\boldXi & \mathbf{S}^{-1}\\
	\end{array} \right] \left[ \begin{array}{c}
		\mathbf{X}' \\
		\mathbf{B}'\\
	\end{array} \right]
	\mathbf{R}_{\phi_w}^{1/2} \mathbf{R}_{\phi_x}^{-1/2} \mathbf{x}\,,
\end{equation}
where $ \mathbf{Q} = (\mathbf{\Tilde{X}'\Tilde{X}})^{-1} + \mathbf{T B' \Tilde{X}} (\mathbf{\Tilde{X}'\Tilde{X}})^{-1} $, $ \mathbf{T} = (\mathbf{\Tilde{X}'\Tilde{X}})^{-1} \mathbf{\Tilde{X}' B S}^{-1} $, $\boldXi = -\mathbf{S}^{-1} \mathbf{B' \Tilde{X}}(\mathbf{\Tilde{X}'\Tilde{X}})^{-1} $, and $ \mathbf{S} = \mathbf{B'B - B' \Tilde{X}} (\mathbf{\Tilde{X}'\Tilde{X}})^{-1} \mathbf{\Tilde{X}' B} $.

We will consider only the first two elements of this expectation vector in what follows. The first term in \eqref{eq:E-theta} gives $\boldbeta$, as expected. The second term refers to the bias, which we indicate as $\Delta_{adj}$, and is equal to:
\begin{equation*}
	\Delta_{adj} = \delta \frac{\sigma_w}{\sigma_z} (\mathbf{Q \Tilde{X}}' - \mathbf{T B}') \mathbf{R}_{\phi_w}^{1/2} \mathbf{R}_{\phi_z}^{-1/2} \mathbf{x}\,.
\end{equation*}

With some algebra, it turns out that
\begin{align*}
	\Delta_{adj} &= \delta \frac{\sigma_w}{\sigma_z} \left\lbrace (\mathbf{\Tilde{X}'\Tilde{X}})^{-1} \mathbf{\Tilde{X}}' + \mathbf{T B' \Tilde{X}} (\mathbf{\Tilde{X}'\Tilde{X}})^{-1} \mathbf{\Tilde{X}}' - \mathbf{T B'} \right\rbrace \mathbf{R}_{\phi_w}^{1/2} \mathbf{R}_{\phi_z}^{-1/2} \mathbf{x}\\
	&= \delta \frac{\sigma_w}{\sigma_z} (\mathbf{\Tilde{X}'\Tilde{X}})^{-1} \mathbf{\Tilde{X}}' \mathbf{R}_{\phi_w}^{1/2} \mathbf{R}_{\phi_z}^{-1/2} \mathbf{x} + \delta \frac{\sigma_w}{\sigma_z} \left\lbrace \mathbf{T B' \Tilde{X}} (\mathbf{\Tilde{X}'\Tilde{X}})^{-1} \mathbf{\Tilde{X}}' - \mathbf{T B'} \right\rbrace \mathbf{R}_{\phi_w}^{1/2} \mathbf{R}_{\phi_z}^{-1/2} \mathbf{x}\\
	&= \Delta_{OLS} + \delta \frac{\sigma_w}{\sigma_z} \mathbf{T B'} \left\lbrace \mathbf{\Tilde{X}} (\mathbf{\Tilde{X}'\Tilde{X}})^{-1} \mathbf{\Tilde{X}}' - \mathbf{I}_n \right\rbrace \mathbf{R}_{\phi_w}^{1/2} \mathbf{R}_{\phi_x}^{-1/2} \mathbf{x}\\
	&= \Delta_{OLS} + \left[\begin{array}{c} d_0 \\ d_x \end{array}\right]
\end{align*}

We have discussed three types of ``null space'', namely:
\begin{enumerate}[label=\emph{Type \arabic*}., itemindent=1cm]
	\setlength\itemsep{1em}
	\item $ \mathbf{U} = \begin{bmatrix} 1 & \mathbf{s}_1[1] & \mathbf{s}_1[2] \\
	& \vdots & \\
	1 & \mathbf{s}_n[1] & \mathbf{s}_n[2] \end{bmatrix} $ and $q=3$
	\item $ \mathbf{U} = \begin{bmatrix} 1 & X(\mathbf{s}_1) & \mathbf{s}_1[1] & \mathbf{s}_1[2] \\
		& \vdots & \vdots & \\
		1 & X(\mathbf{s}_n) & \mathbf{s}_n[1] & \mathbf{s}_n[2] \end{bmatrix} $ and $q=4$
	\item $ \mathbf{U} = \mathbf{\Tilde{X}} $ and $q=2$
\end{enumerate}

Let $ \mathbf{C} $ denote the matrix collecting easting and northing coordinates of the $n$ sampling locations, and $ \mathbf{V}^* $ be the matrix containing the eigenvectors $\mathbf{v}_{q+1}, \dots, \mathbf{v}_n$ corresponding to the non-null eigenvalues.
Under the type-2 null space, it follows that $ \mathbf{A} = \left[ \begin{array}{ccc} \Tilde{\mathbf{X}} & \mathbf{C} & \mathbf{V}^* \end{array} \right] $, and $ \mathbf{B} = \mathbf{V}^* $.

The components of the block inverse in Equation (\ref{eq:blkmat1}) become:
\begin{align*}\arraycolsep=5pt
	\mathbf{S} &= \mathbf{I}_{n-q}, \quad q=4\\
	\mathbf{T} &= \mathbf{0} \\
	\mathbf{Q} &= \left[ \begin{array}{cc}
		\mathbf{\Tilde{X}'\Tilde{X}} & \mathbf{\Tilde{X}'C} \\
		\mathbf{C'\Tilde{X}} & \mathbf{C'C} \\
	\end{array} \right]
\end{align*}

Because of the orthogonality between $ \left[ \begin{array}{cc} \Tilde{\mathbf{X}} & \mathbf{C} \end{array} \right] $ and $ \mathbf{V}^* $, it turns out that $ \Delta_{adj} $ does not depend on the bases in $\mathbf{V^*}$ anymore:
\begin{align*}
	\Delta_{adj} &= \Delta_{OLS} \\
	&+ \delta \frac{\sigma_w}{\sigma_z} \left[ (\mathbf{\Tilde{X}'\Tilde{X}})^{-1} \mathbf{\Tilde{X}}' \mathbf{C} \left\lbrace \mathbf{C}' (\mathbf{I}_n - \mathbf{\Tilde{X}} (\mathbf{\Tilde{X}'\Tilde{X}})^{-1} \mathbf{\Tilde{X}'}) \mathbf{C} \right\rbrace ^{-1} \mathbf{C}' \mathbf{\Tilde{X}} (\mathbf{\Tilde{X}'\Tilde{X}})^{-1} \mathbf{\Tilde{X}}' \right. \\
	&\left. - \left\lbrace \mathbf{X}' (\mathbf{I}_n - \mathbf{C} (\mathbf{C' C})^{-1} \mathbf{C'}) \mathbf{\Tilde{X}} \right\rbrace^{-1} \mathbf{\Tilde{X}}' \mathbf{C} (\mathbf{C' C})^{-1} \mathbf{C'} \right] \mathbf{R}_{\phi_w}^{1/2} \mathbf{R}_{\phi_z}^{-1/2} \mathbf{x}\,,
\end{align*}
so the confounding bias can be reduced only through the inclusion of the spatial coordinates in the model.\\

In the case of type-3 null space, it results that  $ \mathbf{A} = \left[ \begin{array}{cc} \Tilde{\mathbf{X}} & \mathbf{V}^* \end{array} \right] $, $ \mathbf{B} = \mathbf{V}^* $ and that
\begin{align*}
	\mathbf{S} &= \mathbf{I}_{n-q}, \quad q=2\\
	\mathbf{T} &= \mathbf{0} \\
	\mathbf{Q} &= (\mathbf{\Tilde{X}'\Tilde{X}})^{-1}
\end{align*}

Thus, $ \Delta_{adj} = \Delta_{OLS} $ and there is no confounding adjustment.

\section{Local Spike-and-Slab Priors} \label{sec:supp-george-nmig}

In addition to the product moment (pMOM) priors, we consider two additional spike-and-slab prior structures, defining different hierarchical models as described next.

\paragraph{Spike-and-slab with fixed-variance model (SS\_fv)}
Following \citet{george1997approaches}, it is assumed that the $j$-th element of $\widetilde{\boldxi}$ follows a prior distribution given by a mixture of normals, namely:
\begin{align*}
	\xi_j|\gamma_j, \psi_j^2 &\stackrel{ind}{\sim} \gamma_j N(0, \psi_j^2) + (1-\gamma_j) N(0, c_0 \psi_j^2) \,, \quad \mbox{for } j=1, \dots, p-1 \\
	\gamma_j|w &\sim Ber(w)\,,
\end{align*}
where $\gamma_j$ is an indicator variable, $\psi_j^2$ denotes the fixed prior variance assigned to the slab components, and $c_0$ is a small number such that $c_0 \psi_j^2$ defines the prior variance for the spike components. The probability of a large prior variance being assigned to a basis coefficient is assumed fixed, $ w=0.5 $, but it is also possible to consider a beta prior distribution such that $ w \sim Beta(a_w,b_w) $ with usually $ a_w = b_w=1 $ \citep{ishwaran2005spike}.

For the fixed effects ($\beta_0$ and $\beta_x$) we assign independent, zero mean, normal prior distribution with known variance $V_\beta$, and the for the nugget effect ($\widetilde{\sigma}_\epsilon^2$) we assign an inverse-gamma prior with known parameters $a$ and $b$. Hence, the parameter vector containing all the unknowns is $\bm{\Theta} = (\boldbeta', \widetilde{\boldxi}', \widetilde{\sigma}_\epsilon^2, \{\gamma_j\}_{j=1}^{p-1})$.

\paragraph{Spike-and-slab with normal mixture of inverse gamma model (SS\_nmig)}
Another option considered is based on a normal mixture of inverse gamma (NMIG) priors. This allows for more flexibility, since a prior distribution for $ \psi_j^2 $ is introduced as an additional stage in the hierarchy, such that $\psi_j^2 \sim IG(a_\psi, b_\psi)$ \citep{ishwaran2005spike, scheipl2012spike}, and the unknown parameter vector becomes $\bm{\Theta} = (\boldbeta', \widetilde{\boldxi}', \widetilde{\sigma}_\epsilon^2, \{\gamma_j\}_{j=1}^{p-1}, \{\psi_j^2\}_{j=1}^{p-1})$.

The following values are adopted for the hyperparameters in the model: $ V_\beta = 1 $, $ a=2 $, $b=0.1$, $ c_0=10^{-4} $. We also set $ \psi_j^2 = 1 $ for FV priors, or $ a_\psi=2 $ and $ b_\psi=1 $ for NMIG priors.

Finally, as in the main text, note that the basis matrix is not truncated, so $p=n$.

\subsection{Derivation of the Full Conditional Distributions} \label{sec:supp-full-conditional}
The Gibbs sampler \citep{gelfand2000gibbs} is implemented using the following full conditional distributions. Starting from Equation (\ref{eq:semipar_compact2}), the prior variance of the regression coefficients, $ \mathbf{V}_{\boldtheta} $, is a diagonal matrix with first two elements equal to $ V_\beta $, and $ (j+2) $th element given by $ Var\left[\xi_j|\gamma_j,\psi_j^2\right] $. The symbol $ \bullet $ denotes data and all the parameters except for the parameter that is being updated.

\begin{align*}
	\gamma_j|\bullet &\sim Ber\left( \frac{m_{1j}}{m_{1j}+m_{0j}} \right) \,, \quad \mbox{for } j=1, \dots, n \\
	\psi_j^2|\bullet &\sim IG\left( a_\psi + \frac{1}{2}, b_\psi + \frac{\xi_j^2}{2 \{\gamma_j + (1-\gamma_j)c_0\} } \right) \mbox{ (only for NMIG priors)} \\
	\widetilde{\sigma}_\epsilon^2|\bullet &\sim IG\left( a + \frac{n}{2}, b + \frac{1}{2} (\mathbf{y} - \mathbf{A} \boldtheta)' (\mathbf{y} - \mathbf{A} \boldtheta) \right)\\
	\boldtheta |\bullet &\sim N\left( \widetilde{\sigma}_\epsilon^{-2} \mathbf{F} \mathbf{A}^{\prime} \mathbf{y}, \mathbf{F} \right)\,,
\end{align*}
where $ \mathbf{F}^{-1} = \left( \widetilde{\sigma}_\epsilon^{-2} \mathbf{A}^{\prime} \mathbf{A} + \mathbf{V}_{\boldtheta}^{-1} \right) $ and
\begin{align*}
	m_{1j} &\propto w \exp \left(-\frac{\xi_j^2}{2 \psi_j^2} \right)\,,\\
	m_{0j} &\propto (1-w) \frac{1}{\sqrt{c_0}} \exp \left(-\frac{\xi_j^2}{2 c_0 \psi_j^2} \right)\,.
\end{align*}

\section{Additional Findings from the Simulation Study} \label{sec:supp-results-sim}

Some additional results from the simulation study are presented in this section for three representative configurations. The main goal is to provide a more detailed comparison among the competing models. The configurations are chosen based on three combinations of range parameters, namely $\phi_x=0.05, \phi_w=0.5$ (we refer to this combination as \textit{Configuration I}), $\phi_x=0.5, \phi_w=0.05$ (\textit{Configuration II}), and $\phi_x=0.2, \phi_w=0.2$ (\textit{Configuration III}).

\begin{figure}
	\centering
	\includegraphics[width=14cm]{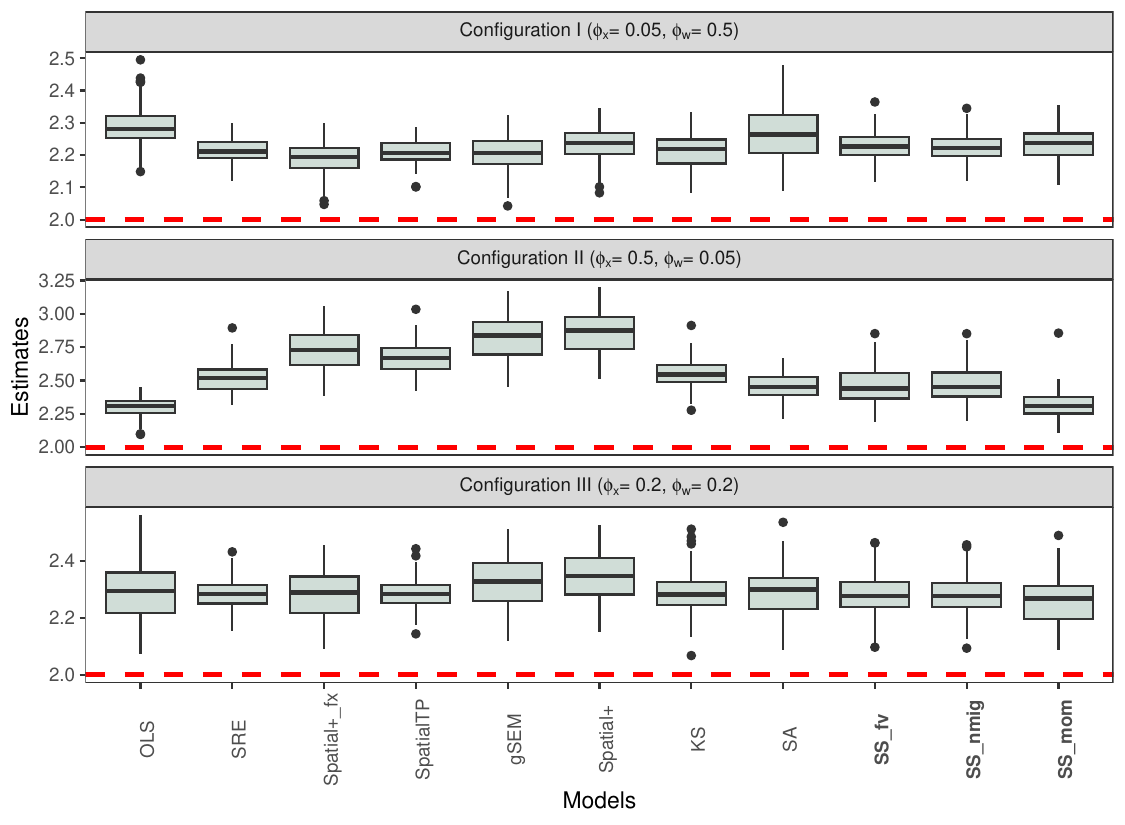}
	\caption[Boxplots for each Configuration]{Boxplots representing the estimated exposure effect for the models in the simulation study, under three configurations where different values are assigned to the range parameters, $\phi_x$ and $\phi_w$. The red dashed line represents the real value, $\beta_x=2$.}
	\label{fig:simulationboxplot}
\end{figure}

The top panel of Figure \ref{fig:simulationboxplot} illustrates the most common scenario in practice where the exposure is less smooth than the unmeasured confounder (\textit{Configuration I}). Whereas Spatial+\_fx provides estimates that are on average the closest to the true value for $\beta_x$ (which is depicted as a red dashed line), all three versions of SS are able to reduce the bias similarly. In the middle panel (\textit{Configuration II}), the confounding bias is usually amplified, however SS\_mom nearly reproduces the non-spatial model's estimates. This may be due to the regularization and model selection scheme implied by the pMOM priors. Indeed, the median number of basis  selected with a posterior inclusion probability greater than $0.5$ is zero in this case. Furthermore, for each one of the 100 replicates, SS\_mom produces estimates that are closer to the true simulated $\beta_x$ than those obtained by the competing approaches. In particular, Spatial+\_fx, (which performed best in Configuration I) always amplifies the bias more than SS\_mom.
In the bottom panel (\textit{Configuration III}), it seems that all approaches except gSEM and Spatial+ are able to reduce the bias on average, but only by a very slight amount. 

\subsection{Difference Between Unadjusted and Adjusted OLS Estimators}

\begin{figure}[ht]
	\centering
	\includegraphics[width=14cm]{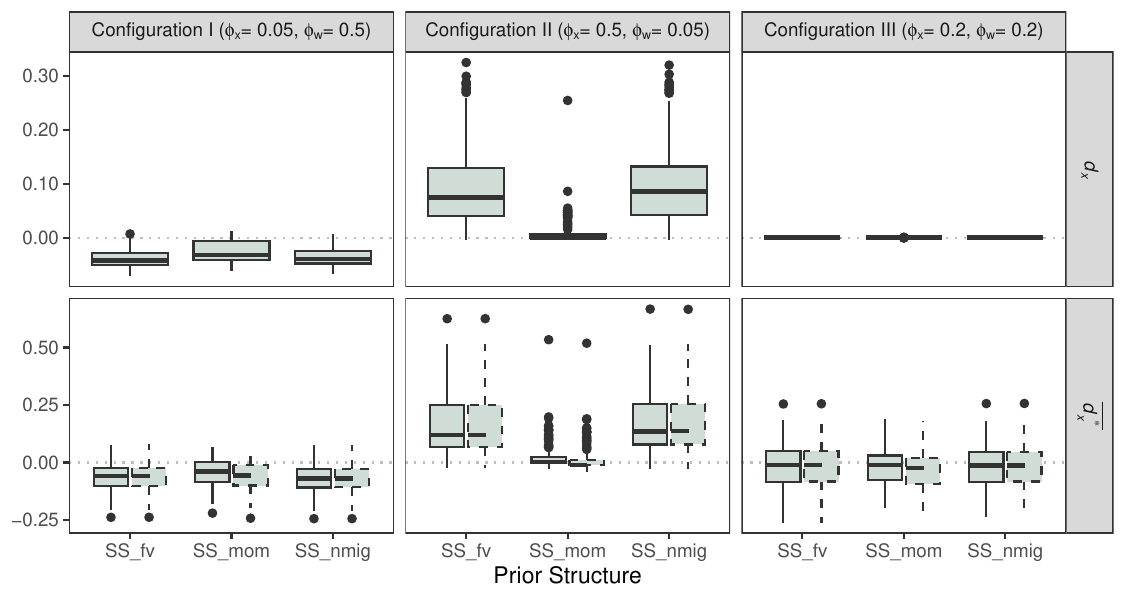}
	\caption[Differences $d_x$ and $d_x^*$: boxplots of \textit{a posteriori} analysis]{For each configuration and for each prior structure, boxplots of \textit{a posteriori} analysis of the amount of bias reduction or amplification are depicted, given the bases selected after each MCMC step. The top panels depict $d_x$, while the bottom panels $\overline{d_x^*}$. The latter is calculated using both analytic (solid lines) and empirical (dashed lines) formulae. Note that the scales on the y-axes are different for each ``row".}
	\label{fig:difference-d-dstar}
\end{figure}

In Section 3.2 of the main manuscript, we have analyzed how $d_x$ changed by progressively increasing the number of bases, $k$, but it is possible to repeat the experiments by considering the covariates selected during each MCMC iteration with posterior inclusion probability $> 0.5$. For each configuration, boxplots of the values of $d_x$ over the 100 replicates are represented in the top row of Figure \ref{fig:difference-d-dstar} for the three types of spike-and-slab priors. Note that the values of $d_x$ may not exactly match the bias reduction or amplification from the corresponding configuration, because the regularization induced by the priors was not taken into consideration in Section 4 of the paper. Nonetheless, using the pMOM priors we generally obtain values of $d_x$ that are negative or close to zero in the worst scenarios: this is because no bases are usually selected when there is risk to amplify the confounding bias.

\subsection{Difference Between Bayesian and Frequentist Estimators}

The foregoing analysis has not taken into consideration how the prior distributions assumed on the basis coefficients impact on the confounding bias. Following a reasoning similar in spirit to that of \citet{hans2009bayesian}, it is possible to compare the unadjusted OLS estimator with the mean of the posterior distribution of $ \beta_x $, when the bases are included. For notational convenience, let the latter quantity be denoted as $\boldbeta_{post} = E[\boldbeta|\bullet]$. We demonstrate that their difference, $d_x^*= \boldbeta_{post} - \widehat{\boldbeta}_{OLS}$, is equal to the second element of the following vector:
\begin{equation}\label{eq:d_star}
	\left[\begin{array}{c} d_0^* \\ d_x^* \end{array}\right] = \mathbf{T B'} (\widehat{\mathbf{Y}}_{OLS} - \mathbf{Y})/\widetilde{\sigma}_\epsilon^2\,,
\end{equation}
where $\widehat{\boldbeta}_{OLS}$ and $\widehat{\mathbf{Y}}_{OLS} = \Tilde{\mathbf{X}} \widehat{\boldbeta}_{OLS} $ are, respectively, the estimate and the vector of fitted values under the unadjusted OLS model, but now $\mathbf{T}$ is not the same as before since it derives from the inversion of another block matrix which depends on the priors.

\begin{proof}
	Starting again from Equation (\ref{eq:semipar_compact2}), now the expected value of the conditional posterior of $ \boldtheta $ is of interest. If non-informative independent priors are assumed on all the parameters, we have:
	\begin{equation*}
		E \left[ \boldtheta | \bullet \right] = \left( \mathbf{A'A}/\widetilde{\sigma}_\epsilon^2 + \mathbf{V}_{\boldtheta}^{-1} \right)^{-1} \mathbf{A'y}/\widetilde{\sigma}_\epsilon^2
	\end{equation*}
	where
	\begin{equation*}
		\mathbf{V}_{\boldtheta}^{-1} = \left[ \begin{array}{cc}
			\mathbf{V}_{\boldbeta}^{-1} & \mathbf{0} \\
			\mathbf{0} & \mathbf{V}_{\boldxi}^{-1}\\
		\end{array} \right]\,.
	\end{equation*}

	Since we use non-informative priors on $ \boldbeta $, the upper-left block of $ \mathbf{V}_{\boldtheta}^{-1} $ tends to the null matrix, thus:
\begin{align*}
	\left( \mathbf{A'A}/\widetilde{\sigma}_\epsilon^2 + \mathbf{V}_{\boldtheta}^{-1} \right)^{-1} &\approx \arraycolsep=5pt \left[ \begin{array}{cc}
		\mathbf{\Tilde{X}'\Tilde{X}}/\widetilde{\sigma}_\epsilon^2 & \mathbf{\Tilde{X}' B}/\widetilde{\sigma}_\epsilon^2 \\
		\mathbf{B' \Tilde{X}}/\widetilde{\sigma}_\epsilon^2 & \mathbf{B'B}/\widetilde{\sigma}_\epsilon^2 + \mathbf{V}_{\boldxi}^{-1}\\
	\end{array} \right]^{-1} = \\
	& = \left[ \begin{array}{cc}
		\mathbf{Q} & -\mathbf{T} \\
		\boldXi & \mathbf{S}^{-1}\\
	\end{array} \right]
\end{align*}
where
\begin{gather*}
	\mathbf{S} = \mathbf{B'B}/\widetilde{\sigma}_\epsilon^2 + \mathbf{V}_{\boldxi}^{-1} - \mathbf{B}' \mathbf{\Tilde{X}} (\mathbf{\Tilde{X}'\Tilde{X}})^{-1} \mathbf{\Tilde{X}' B}/\widetilde{\sigma}_\epsilon^2\,, \\
	\mathbf{T} = (\mathbf{\Tilde{X}'\Tilde{X}})^{-1} \mathbf{\Tilde{X}' B S}^{-1}\,, \\
	\mathbf{Q} = (\mathbf{\Tilde{X}'\Tilde{X}})^{-1}\widetilde{\sigma}_\epsilon^2 + \mathbf{T B' \Tilde{X}} (\mathbf{\Tilde{X}'\Tilde{X}})^{-1}\,, \\
	\boldXi = -\mathbf{S}^{-1} \mathbf{B' \Tilde{X}}(\mathbf{\Tilde{X}'\Tilde{X}})^{-1}\,.
\end{gather*}

Now we can focus on the first two elements of $ \boldtheta $, i.e., $ \boldbeta $, to obtain:
\begin{align*}
	\boldbeta_{post} &= (\mathbf{Q X' - T B'}) \mathbf{y}/\widetilde{\sigma}_\epsilon^2 \\
	&= \widehat{\boldbeta}_{OLS} + \mathbf{T B' \Tilde{X}} \widehat{\boldbeta}_{OLS}/\widetilde{\sigma}_\epsilon^2 - \mathbf{T B' Y}/\widetilde{\sigma}_\epsilon^2\\
	&= \widehat{\boldbeta}_{OLS} + \mathbf{T B'} (\widehat{\mathbf{Y}}_{OLS} - \mathbf{Y})/\widetilde{\sigma}_\epsilon^2 \\
	&= \widehat{\boldbeta}_{OLS} + \left[\begin{array}{c} d_0^* \\ d_x^* \end{array}\right]\,.
\end{align*}

\end{proof}

Therefore, $d_x^*$ can be obtained using either an ``empirical formula", that is as a difference between Bayesian and frequentist estimators, or an ``analytic formula", which depends on the bases, the residuals from the unadjusted OLS fit, and the prior structure assumed on the basis coefficients.\\

For the type-2 null space, we are not assuming spike-and-slab priors on the spatial coordinates, so a large variance is imposed to the respective coefficients, and the previous formulae can be simplified: from the orthogonality between $ \left[ \begin{array}{cc} \Tilde{\mathbf{X}} & \mathbf{C} \end{array} \right] $ and $ \mathbf{V}^* $, the matrix $\left( \mathbf{A'A}/\widetilde{\sigma}_\epsilon^2 + \mathbf{V}_{\boldtheta}^{-1} \right) $ becomes block-diagonal and its inverse is:
\begin{equation*}
	\left( \mathbf{A'A}/\widetilde{\sigma}_\epsilon^2 + \mathbf{V}_{\boldtheta}^{-1} \right)^{-1} \approx \arraycolsep=5pt \left[ \begin{array}{cc}
		\widetilde{\sigma}_\epsilon^2 \left[ \begin{array}{cc} \Tilde{\mathbf{X}}' \Tilde{\mathbf{X}} & \mathbf{\Tilde{X}' C} \\
			\mathbf{C' \Tilde{X}} & \mathbf{C' C} \end{array} \right]^{-1} & \mathbf{0} \\
		\mathbf{0} & \left(\mathbf{I}_{n-q}/\widetilde{\sigma}_\epsilon^2 + \mathbf{V^*}_{\boldxi}^{-1} \right)^{-1}\\
	\end{array} \right]\,,
\end{equation*}
where $\mathbf{V^*}_{\boldxi}$ is $\mathbf{V}_{\boldxi}$ without the first two columns and the first two rows.

It turns out that the posterior mean does not depend on the bases:
\begin{align*}
	\boldbeta_{post} &= \widehat{\boldbeta}_{OLS} + (\mathbf{\Tilde{X}'\Tilde{X}})^{-1} \mathbf{\Tilde{X}}' \mathbf{C} \left\lbrace \mathbf{C}' (\mathbf{I}_n - \mathbf{\Tilde{X}} (\mathbf{\Tilde{X}'\Tilde{X}})^{-1} \mathbf{\Tilde{X}'}) \mathbf{C} \right\rbrace ^{-1} \mathbf{C'} \widehat{\mathbf{Y}}_{OLS} \\
	&- \left\lbrace \mathbf{X}' (\mathbf{I}_n - \mathbf{C} (\mathbf{C' C})^{-1} \mathbf{C'}) \mathbf{\Tilde{X}} \right\rbrace ^{-1} \mathbf{\Tilde{X}}' \mathbf{C} (\mathbf{C' C})^{-1} \mathbf{\Tilde{X}'} \mathbf{Y} \,,
\end{align*}

As for the type-3 null space, given the independence between $ \mathbf{\Tilde{X}} $ and each eigenvector, the posterior mean simplifies to the OLS estimate. In fact:
\begin{align*}
	\left( \mathbf{A'A}/\widetilde{\sigma}_\epsilon^2 + \mathbf{V}_{\boldtheta}^{-1} \right)^{-1} &\approx \arraycolsep=5pt \left[ \begin{array}{cc}
		\mathbf{\Tilde{X}'\Tilde{X}}/\widetilde{\sigma}_\epsilon^2 & \mathbf{0} \\
		\mathbf{0} & \mathbf{I}_{n-q}/\widetilde{\sigma}_\epsilon^2 + \mathbf{V}_{\boldxi}^{-1}\\
	\end{array} \right]^{-1} = \\
	& = \left[ \begin{array}{cc}
		(\mathbf{\Tilde{X}'\Tilde{X}})^{-1} \widetilde{\sigma}_\epsilon^2 & \mathbf{0} \\
		\mathbf{0} & (\mathbf{I}_{n-q}/\widetilde{\sigma}_\epsilon^2 + \mathbf{V}_{\boldxi}^{-1})^{-1}\\
	\end{array} \right]
\end{align*}
which implies that $\boldbeta_{post} = (\mathbf{\Tilde{X}'\Tilde{X}})^{-1} \mathbf{\Tilde{X}' Y} = \widehat{\boldbeta}_{OLS}$.\\

The bottom panels of Figure \ref{fig:difference-d-dstar} show boxplots of $\overline{d_x^*} = E[d_x^*|\bullet]$ over the hundred replicates. The results, using both empirical (represented by dashed lines) and analytic (solid lines) formulae, are plotted for each configuration and prior structure. There is an overall correspondence of the results obtained by the formulae, however in the pMOM case a workaround was necessary. Since there is not a closed-form full conditional distribution for the coefficients, it was not possible to derive an analytic formula. Hence, we show what would have happened if the bases in the pMOM case were selected under the fixed-variance prior structure. Again, we see that the results are consistent with our findings in Section 5 of the paper, and that SS\_mom performs better than the other two approaches, since $\overline{d_x^*}$ is almost always smaller than or equal to zero.\\

\subsection{Bases Selected by the Competing Approaches}

An important question relates to the number of bases selected by the competing models. This section provides a summary of the results, which are presented in Table \ref{sim-results-EDF}, for the competing low-rank models. As for SpatialTP ans Spatial+, the maximal degrees of freedom prior to penalization is $df = 150$. For KS, the degrees of freedom of the (unpenalized) spline term are determined using AIC, whereas the bases selected by the SS models are determined by the prior structure.

Since the SpatialTP and Spatial+ models are generalized additive models \citep[GAMs;][]{wood2017generalized} and their spline terms receive a penalization to encourage smoothness, the effective degrees of freedom (EDF) are usually smaller than the maximal degrees of freedom. For these models, Table \ref{sim-results-EDF} reports the median EDF over the 100 replicates for each configuration. Spatial+ estimates a higher EDF than SpatialTP in all cases, probably because the smooth term has to account for the spatial dependence not only of the unmeasured confounder but also of the exposure. In addition, conversely to SpatialTP, the EDF for Spatial+ is the smallest in Configuration II, but it is not small enough to avoid confounding bias amplification.

As for the KS approach, with EDF we mean the number of bases selected by minimizing the AIC. In Configuration I, the median EDF equals the maximum allowed and is the highest among all models considered so far, but this does not provide it with an advantage in terms of bias reduction (see Figure \ref{fig:simulationboxplot}).

In the case of the SA model, the EDF is the number of bases selected by minimizing the deviance information criterion (DIC). Finally, the EDF for the SS models are defined as the number of bases with posterior inclusion probability $>50\%$. 

Whereas SpatialTP, Spatial+, and KS use a basis expansion of TPRS, the SA and SS models use a basis expansion of B-splines and of principal splines, respectively. Therefore, the results are not directly comparable. However, we can still compare the EDFs among the SS models. The SS\_mom approach is the most parsimonious, with the lowest median EDF in all configurations. Among the configurations, all the SS models have the highest EDF in Configuration I, and select the least number of bases in Configuration II, corroborating the results in Figure \ref{fig:simulationboxplot}. Notably, the median EDF of SS\_mom is zero in Configuration II.

\begin{table}[!htb]
	\centering
	\caption{Median EDF of the smooth term under different models and for each configuration.}
	\label{sim-results-EDF}
	\begin{tabular}{cccccccccccccccc}
		\Hline
		~ &  SpatialTP & Spatial+ & KS & SA & \textbf{SS\_fv} & \textbf{SS\_nmig} & \textbf{SS\_mom} \\ 
		\hline
		Configuration I & 75.56 & 138.42 & 250 & 20 & 16 & 13 & 7 \\ 
		Configuration II & 71.66 & 126.39 & 30 & 12 & 9 & 9 & 0 \\ 
		Configuration III & 58.61 & 136.58 & 30 & 12 & 11 & 9 & 3 \\ 
		\hline
	\end{tabular}
\end{table}

\begin{figure}[ht]
	\centering
	\includegraphics[width=13cm]{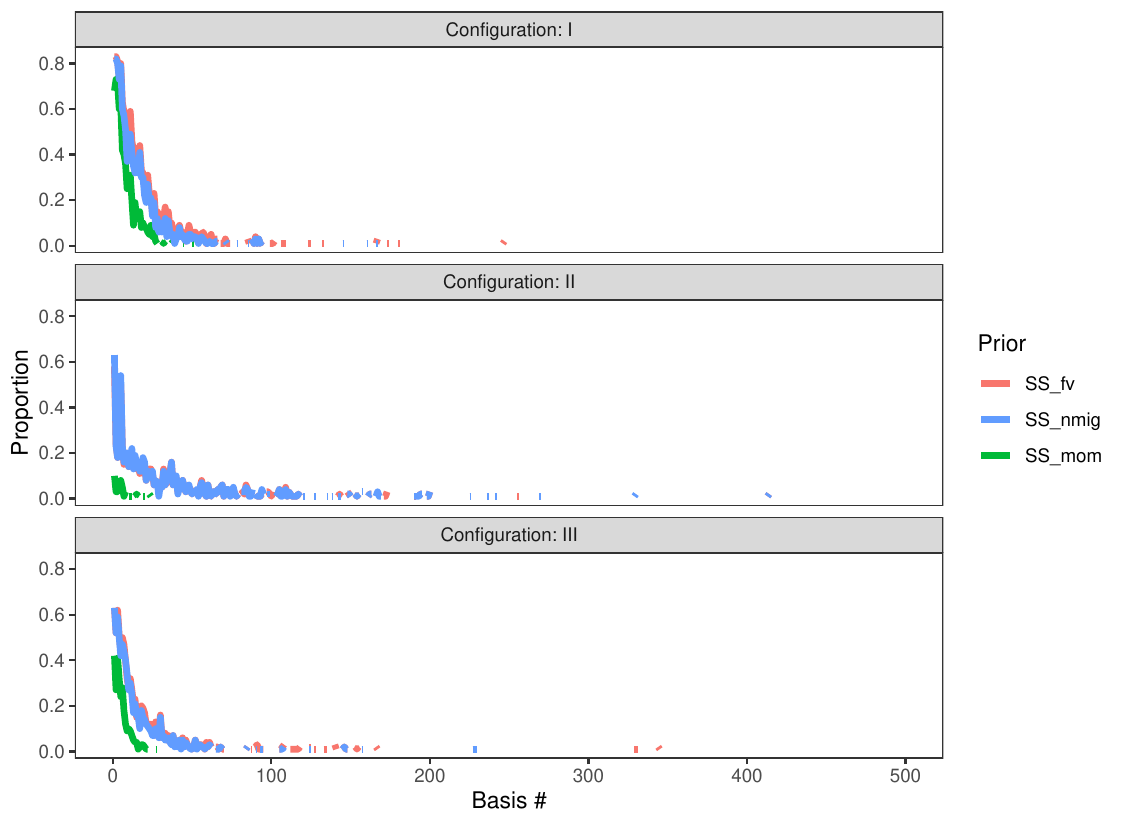}
	\caption{Proportion of each basis being selected by the SS models with posterior inclusion probability $>50\%$, for the three configurations.}
	\label{fig:proportion-bases-selected}
\end{figure}

It is possible to investigate how the non-linear selection of bases works. Figure \ref{fig:proportion-bases-selected} depicts the proportion of each basis being selected by the SS models with posterior inclusion probability $>50\%$. Each colored line represents a prior structure, and the three panels refer to the three configurations. The lines are discontinuous because the bases with zero proportion are not plotted. Successive bases are often selected with different proportions, which is a sign of the non-linear selection mechanism. In contrast, the other approaches select the first bases with a constant proportion (equal to 1) and without discontinuities: this reflects the fact that the competing approaches require prior knowledge about the bases to be selected.

\subsection{Other Simulation Scenarios} \label{sec:supp-other-scenarios}
The following Table \ref{more-results-beta} reports the results of the simulation study for other configurations not considered in the main manuscript. When $\delta = 0$, all the models should provide unbiased estimators for $\beta_x$. Unexpectedly, this seems not to be true for gSEM and Spatial+ with estimated smoothing parameter. Then, as $|\delta|$ increases, the confounding bias under any approach increases proportionally. There is an overall symmetry between what happens given negative and positive values for $\delta$, with the exception of  gSEM and Spatial+, which tend to produce higher bias if $\delta$ is positive, but smaller (absolute) bias if $\delta$ is negative, when compared to Spatial+\_fx (which behaves symmetrically). Next, according to all the models being compared, $\sigma_x$ and $\sigma_w$ have linear effects on the bias, as expected.

Finally, the number of bases selected by the SS models depends not only on the relationship between the range parameters. Indeed, if the partial sill of the unmeasured confounder increases, the number of bases considered relevant will consequently increase. The reason could be that there is more residual spatial variability to explain. The prior structure also plays an important role, since pMOM priors produce more parsimonious models than FV or NMIG priors.

\begin{sidewaystable}
	\centering
	\caption{Estimates, averaged over 100 replicates, of $\beta_x$ for some other configurations.}
	\label{more-results-beta}
	\begin{tabular}{cccccccccccccccc}
		\hline
		$\phi_x$ & $\phi_w$ & $\delta$ & $\sigma_x^2$ & Null & SRE & Spatial+\_fx & SpatialTP & gSEM & Spatial+ & KS & SA & \textbf{SS\_fv} & \textbf{SS\_nmig} & \textbf{SS\_mom} \\ 
		\hline
		0.20 & 0.20 & 0 & 1.00 & 2.01 & 2.00 & 2.00 & 2.00 & 2.05 & 2.08 & 1.97 & 2.01 & 1.99 & 2.00 & 1.97 \\ 
		0.50 & 0.05 & 0.2 & 1.00 & 2.30 & 2.55 & 2.76 & 2.72 & 2.86 & 2.93 & 2.51 & 2.53 & 2.51 & 2.52 & 2.37 \\ 
		0.20 & 0.20 & 0.2 & 1.00 & 2.30 & 2.28 & 2.29 & 2.28 & 2.33 & 2.37 & 2.25 & 2.29 & 2.28 & 2.28 & 2.25 \\ 
		0.05 & 0.50 & 0.2 & 1.00 & 2.29 & 2.20 & 2.19 & 2.20 & 2.19 & 2.25 & 2.20 & 2.21 & 2.21 & 2.21 & 2.21 \\ 
		0.50 & 0.05 & -0.5 & 1.00 & 1.71 & 1.51 & 1.33 & 1.37 & 1.42 & 1.45 & 1.46 & 1.56 & 1.55 & 1.54 & 1.66 \\ 
		0.20 & 0.20 & -0.5 & 1.00 & 1.71 & 1.71 & 1.72 & 1.71 & 1.77 & 1.80 & 1.70 & 1.71 & 1.71 & 1.71 & 1.69 \\ 
		0.05 & 0.50 & -0.5 & 1.00 & 1.71 & 1.78 & 1.80 & 1.79 & 1.82 & 1.85 & 1.73 & 1.73 & 1.78 & 1.78 & 1.75 \\ 
		0.50 & 0.05 & 0.5 & 0.50 & 2.29 & 2.47 & 2.63 & 2.59 & 2.77 & 2.82 & 2.51 & 2.39 & 2.41 & 2.41 & 2.30 \\ 
		0.20 & 0.20 & 0.5 & 0.50 & 2.29 & 2.28 & 2.26 & 2.28 & 2.34 & 2.37 & 2.28 & 2.27 & 2.27 & 2.27 & 2.26 \\ 
		0.05 & 0.50 & 0.5 & 0.50 & 2.29 & 2.22 & 2.19 & 2.21 & 2.22 & 2.26 & 2.21 & 2.28 & 2.23 & 2.23 & 2.24 \\ 
		0.50 & 0.05 & 0.5 & 2.00 & 2.30 & 2.57 & 2.80 & 2.73 & 2.86 & 2.89 & 2.57 & 2.52 & 2.54 & 2.56 & 2.35 \\ 
		0.20 & 0.20 & 0.5 & 2.00 & 2.30 & 2.29 & 2.29 & 2.29 & 2.32 & 2.33 & 2.29 & 2.30 & 2.29 & 2.29 & 2.26 \\ 
		0.05 & 0.50 & 0.5 & 2.00 & 2.30 & 2.21 & 2.19 & 2.21 & 2.20 & 2.22 & 2.22 & 2.24 & 2.23 & 2.22 & 2.22 \\
		\hline
	\end{tabular}
\end{sidewaystable}

\begin{figure}[ht]
	\centering
	\includegraphics[width=14cm]{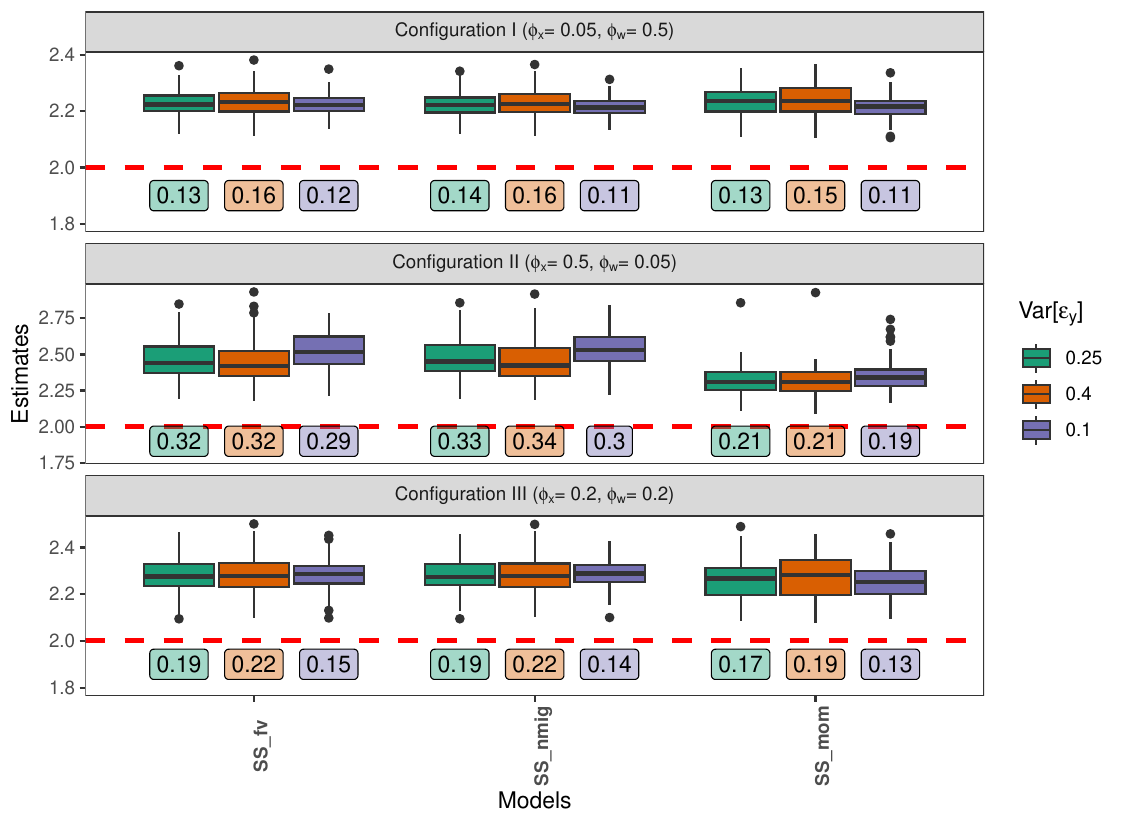}
	\caption[Boxplots for each Configuration]{Boxplots representing the estimated exposure effect under for the proposed model, under the three configurations. The red dashed line represents the real value, $\beta_x=2$. The colors represent the different values of $ \sigmaeps^2 = Var[\epsilon_y(\mathbf{s})] $. The numbers in colored boxes indicate the average width of the 95\% credible intervals.}
	\label{fig:more-results-sigma2eps}
\end{figure}

So far, the variance of the measurement error has been set to $ \sigmaeps^2=0.25 $, but it is interesting to investigate whether the results obtained by the proposed approach change when this parameter is varied. We now consider the values $ \sigmaeps^2 \in \{0.1, 0.25, 0.4 \} $ to change the signal-to-noise ratio. Figure \ref{fig:more-results-sigma2eps} is similar in spirit to Figure \ref{fig:simulationboxplot}, but now there are boxplots of three different colors, each representing a different value of $ \sigmaeps^2 $ (the green ones are those from Figure \ref{fig:simulationboxplot}). In terms of confounding bias, the performance of the proposed model are generally not affected by the choice of $ \sigmaeps^2 $. An exception is in Configuration II, where the estimates of SS\_fv and SS\_nmig slightly increase as $ \sigmaeps^2 $ decreases, while the estimates of SS\_mom remain stable. The widths of the 95\% credible intervals (averaged over the replicates) are also reported in the colored boxes, and they increase when the signal-to-noise ratio decreases, as expected.

\subsection{Simulation with Different Exposures} \label{sec:supp-different-X}

\begin{figure}[ht]
	\centering
	\includegraphics[width=13cm]{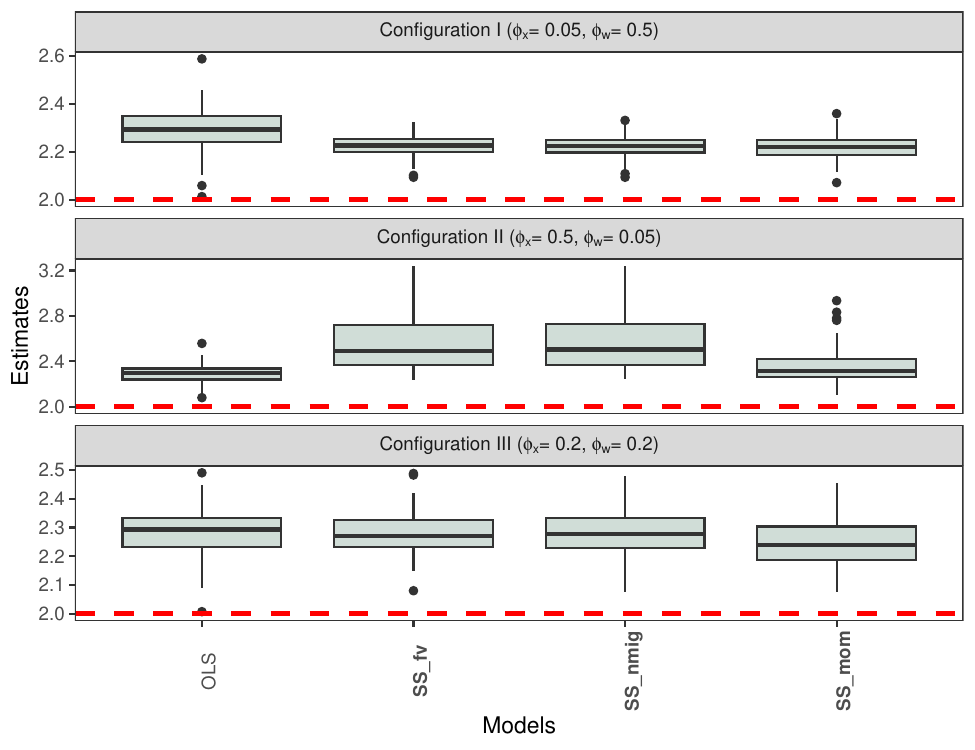}
	\caption[Simulation with Different Exposures]{Boxplots representing the estimated exposure effect for OLS and the proposed models, under the three configurations, when the exposure is not held fixed across replicates. The red dashed line represents the real value, $\beta_x=2$.}
	\label{fig:more-results-different-X}
\end{figure}

The main simulation study assumes that the exposure is held fixed within a given configuration. However, the results discussed so far may be deeply influenced by the sampled exposure. Keeping all other aspects of the setup as in the main manuscript, we now consider a different exposure surface for each replicate. Figure \ref{fig:more-results-different-X} shows that there is not any relevant changes in terms of bias. The only change is in terms of the width of the boxplots, which is slightly larger given the increased variability in the data. Since these results are consistent with the findings in the main manuscript and to avoid redundancies, Figure \ref{fig:more-results-different-X} only reports the estimates for OLS and the variants of the proposed model.

\end{document}